\documentclass[USenglish,oneside,twocolumn]{article}

\usepackage[utf8]{inputenc}
\usepackage[big]{dgruyter_NEW}
\usepackage[nounderscore]{syntax}
\usepackage{mdframed}
\usepackage{xcolor}
\usepackage{amsmath}
\usepackage{amssymb}
\usepackage{caption}
\usepackage{subcaption}
\usepackage{listings}
\usepackage{makecell}

\DOI{foobar}

\cclogo{\includegraphics{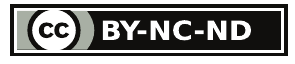}}
  
\begin{document}
 \newcommand{\fixme}[1]{\textcolor{red}{#1}}

   \author*[1]{Shaowei Zhu}
   
   \author[2]{Hyo Jin Kim}
   
   \author[3]{Maurizio Monge}
   
   \author[4]{G. Edward Suh}
   
   \author[5]{Armin Alaghi}
   
   \author[6]{Brandon Reagen}

   \author[7]{Vincent Lee}

   \affil[1]{Facebook Reality Labs Research, E-mail: shaoweiz@fb.com}

   \affil[2]{Facebook Reality Labs Research, E-mail: kimhyojin@fb.com}
   
    \affil[3]{Facebook Reality Labs Research, E-mail: maurimo@fb.com}
    
     \affil[4]{Facebook AI Research and Cornell University, E-mail: edsuh@fb.com}
     
      \affil[5]{Facebook Reality Labs Research, E-mail: alaghi@fb.com}
      
       \affil[6]{Facebook Reality Labs Research and New York University, E-mail: reagen@fb.com}

  \affil[7]{Facebook Reality Labs Research, E-mail: vtlee@fb.com}

  \title{\huge Verifiable Access Control for Augmented Reality Localization and Mapping}

  \runningtitle{Verified Access Control for Localization and Mapping}


  \begin{abstract}
  {
  Localization and mapping is a key technology for 
  bridging the virtual and physical worlds in augmented reality (AR). 
  Localization and mapping works by creating and querying maps made of anchor points that enable the overlay of these two worlds.
  As a result, 
  information about the physical world is captured in the
  map and naturally gives rise to concerns around who can map physical spaces as
  well as who can access or modify the virtual ones.
  This paper discusses how we can
  provide access controls over virtual
  maps as a basic building block to enhance security and privacy of AR systems.
  In particular, we
  propose VACMaps:
  an access control system for localization and mapping
  using formal methods. 
  VACMaps defines a domain-specific language that enables users to specify
  access control policies for virtual spaces.
  Access requests to virtual spaces are then evaluated against relevant policies 
  in a way that preserves confidentiality and integrity of virtual spaces owned by the
  users.
  The precise semantics of the policies are defined by SMT formulas, which allow
  VACMaps to reason about properties of access policies automatically.
  An evaluation of VACMaps is provided using an AR testbed of a single-family home.
  We show that VACMaps is scalable in that it can run at
  practical speeds and that it can also reason about access control policies automatically to detect potential policy misconfigurations.
  }
\end{abstract}



  \journalname{Proceedings on Privacy Enhancing Technologies}
\DOI{Editor to enter DOI}
  \startpage{1}
  \received{..}
  \revised{..}
  \accepted{..}

  \journalyear{..}
  \journalvolume{..}
  \journalissue{..}

\maketitle

\newcommand{\policydslname}{\texttt{VMAC Lang}}

\section{Introduction}
\label{sec:introduction}

Augmented reality (AR) is rapidly emerging as the next generation of disruptive technology.
Like the mobile computing revolution, AR is similarly poised to change how the physical and digital world connect and interact.
It provides immense opportunity for assisting and improving how we interact with the physical world by precisely overlaying just the right amount of relevant information from the virtual one.
Augmented reality promises to provide a wide range of applications from social teleportation~\cite{Saito_2020_CVPR,kato2000virtual}, e-commerce~\cite{han2018viton,lu2007augmented}, navigation~\cite{narzt2006augmented}, and contextualized artificial intelligence~\cite{context-ai1, context-ai2}.

There are countless exciting use cases for AR but all are built upon \emph{localization} and \emph{mapping}.
Localization and mapping is responsible for determining where a user is in physical space and aligning it with the virtual world.
Mapping is the process of sampling (physical) data to build new maps or update existing ones.
These maps are the backbone of AR systems 
that bridge the physical and virtual divide.
A commonly used form of a vision-based virtual map are 3D point clouds, which are composed of points in 3D space and their associated visual feature descriptors that describe the appearance of each point location (\autoref{fig:point_cloud_example}).
This representation contains the information necessary to associate a point in physical space with one in the virtual world.

Once a virtual map equivalent to the physical environment has been constructed, users interact with the map to localize into the environment.
Localization is the process of querying existing maps using sensor inputs to determine where a user is in the environment.
This information is used to resolve the relative locations of virtual and physical assets, enabling a precise overlay of the virtual world on top of the physical one.
For instance, suppose we want to place a virtual sticky note on a physical refrigerator (assuming the physical space has been mapped).
To render the virtual sticky note in the right spot the AR system needs the following information: (1) the physical location of the refrigerator, (2) the virtual location of the sticky note, (3) the location of the user in both physical and virtual space, and (4) the relative distance between the note, refrigerator, and user in both the physical and virtual world. 
Localization and mapping provides all this information by aligning the virtual world to the physical one.

While the virtual AR world has remarkable likeness to the physical world, there are important differences with respect to accessing spaces, which now depend on access to digital maps rather than just physical locations.
This gives rise to the map access problem in AR.
Specifically, unlike the physical world where restricting access is intuitive and solutions are readily available (e.g., locks and physical barriers), new systems and mechanisms are needed for controlling access to virtual spaces for AR.
Access control systems should 
uphold and enforce notions of permission for map modifications, analogous to the physical world.
Imagine two competing coffee shops. 
One shop should not be able to place advertisements for their shop in the line of the other, nor should 
it be able to artificially obstruct/close the entryways in the virtual world.
More questions arise around who, when, and where a map can be used for localization.
For example, if a map exists of a house, how does one enforce who can interact with which parts and when? 
Without restrictions, these digital spaces are ripe to become overrun with unwanted behavior.

In this paper we propose and develop Verifiable Access Controlled Maps (i.e., VACMaps)
to address the emerging challenges of access control in AR.
VACMaps is an access control system that uses formal methods to implement a wide range of configurable access control policies over maps, specifically targeting mapping and localization use cases.
It also maintains a hierarchy of spaces enabling efficient look-up of relevant spaces and policies.
We develop a new domain-specific language, named  
\policydslname{} (Virtual Map Access Control Language) to specify access control policies.
The language enables the owner of a space to detail precisely who should have write (i.e., mapping) and read (i.e., localization) access to a map and under what conditions (e.g., the time of day and a user's current location).
The semantics of this language is given by a set of rules that translate the human-readable policies to
logical formulas.
Once VACMaps receives an access request, it efficiently collects relevant spaces and their corresponding
access policies, and then evaluate the access request against the logical formula representation 
of the policies.
Thanks to the logical formula representation of policies, VACMaps is able to formally prove claims 
such as ``an unauthorized user will not have access to this space'', thereby protecting the victim's private map contents.

VACMaps can verify access claims to map data given a set of access policies but there is still
possibility for users to unintentionally misconfigure the access policies themselves (e.g., overly permissive access, etc.).
While VACMaps would not be able to figure out the user's real intent, it does provide a way to detect
and audit potential misconfigurations.
This is only possible because of the formula representation of policies that VACMaps maintains. 
We can then craft queries questioning VACMaps ``who has access to this room?'' or ``is there a space in this house that no one has access to?'' or ``would this new policy have any effect at all?''
Answers to these queries might be useful for users in order for them to author policies in a 
more informed manner.

This paper makes the following contributions:
\begin{itemize}

    \item We propose VACMaps: the first system that implements access control for modern AR localization and mapping using formal methods. 
    \item We develop \policydslname{}: a domain-specific language for access policies whose semantics are defined by the SMT encodings.
    \item We evaluate VACMaps and show that it can run in near real-time and can scale to large numbers of spaces and access policies.
    \item We demonstrate how VACMaps can be used to audit access rights and detect possible policy
    misconfigurations.
    
\end{itemize}


\section{Background and Motivation}
\label{sec:background}

This section provides background on localization and mapping and motivates the need for map access control.

\begin{figure*}
\centering
\includegraphics[width=\linewidth]{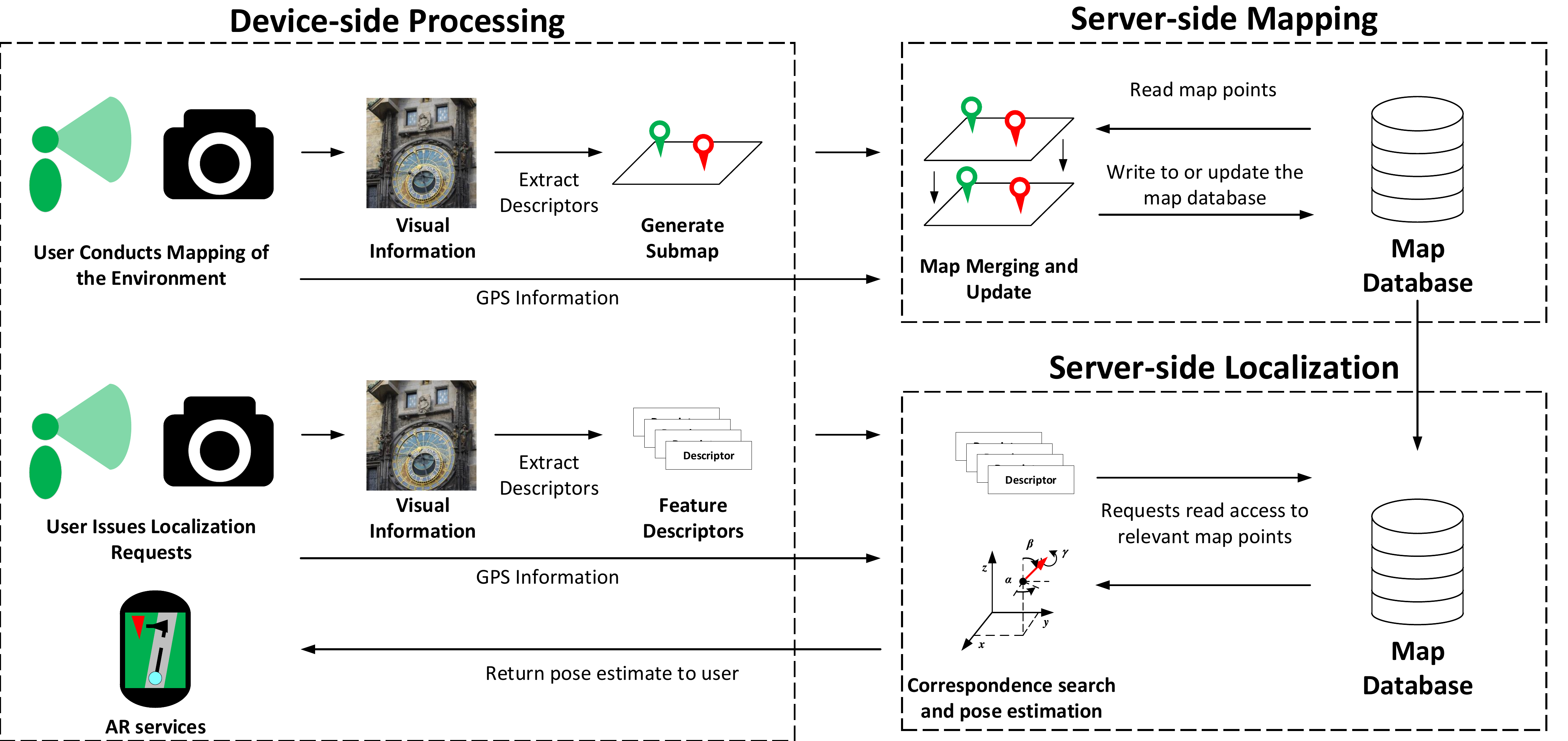}
\caption{Localization and mapping is a two part process. In mapping, virtual 3D maps of physical spaces are created from the user's scanning of the space. These maps are stored on the server and also used to update the existing maps. Coarse-grained GPS location information is used to reduce the map search space. During localization, a set of features are extracted from user's device input stream which are matched against map points to estimate the device's pose in the environment.}
\label{fig:localization_and_mapping}
\end{figure*}

\subsection{Background: Maps, Mapping, and Localization}

Localization and mapping is a two part process as shown in \autoref{fig:localization_and_mapping}.
\textbf{Mapping} is the process of creating a 3D model, usually in the form of 3D point clouds, of a physical space through computer vision algorithms, such as Simultaneous Localization and Mapping (SLAM) \cite{mur-artal2016orb-slam,engel2014lsd,newcombe2011dtam} or Structure-from-Motion (SfM) \cite{schonberger2016colmap,sweeney2015theia,agarwal2011building}.
The mapping process typically takes an image from the AR device as input and generates a visual feature descriptor for each interest point detected in the image.
Each interest point is also associated with a location in physical 3D space (i.e., x, y, and z coordinates).
These map points are then combined to create \textbf{maps}---the core data structure used to represent the physical world in machine perception algorithms.
The mapping process effectively provides a binding between locations in a physical space and the digital copy for AR, which is encoded in the map. 
We refer to this digital copy of physical space encoded in map data as \textit{virtual spaces}.
Each new map then is merged into the map database on the server where GPS data is used to provide a coarse grained estimate for roughly where to align and merge the new map with.

\textbf{Localization} is responsible for using map data and obtaining a high-precision estimate of where a user is in both the physical and virtual world.
Localization provides the 6 degree-of-freedom (DoF) pose of the user's device by querying an existing map.
This is usually achieved by registering a device's input video stream against the map (3D model) of the physical space; additional signals, such as GPS, are also used to get a coarse-grained position estimate in the world to reduce the search space~\cite{sattler2017large,zeisl2015camera,chen2011city}.
Specifically, localization uses a set of visual descriptors extracted from user device's input video stream and matches them against the descriptors that are associated with each 3D point in the map. 
Without a coarse grained GPS estimate, this matching step would search significantly larger portions of the database to get high quality matches for localization.
Once the correspondences have been established, the pose is computed via a PnP algorithm~\cite{kukelova2013real,bujnak2010new,haralick1994review}.
In this work, we assume that localization and mapping are performed on the server. 
In this scenario, the user sends visual information (e.g., visual descriptors) for localization, the server performs localization and then sends the 6-DoF pose back to the user. 

\subsection{Access Control Considerations for Virtual Spaces}

\begin{figure*}[t]
\centering
\includegraphics[width=0.8\linewidth]{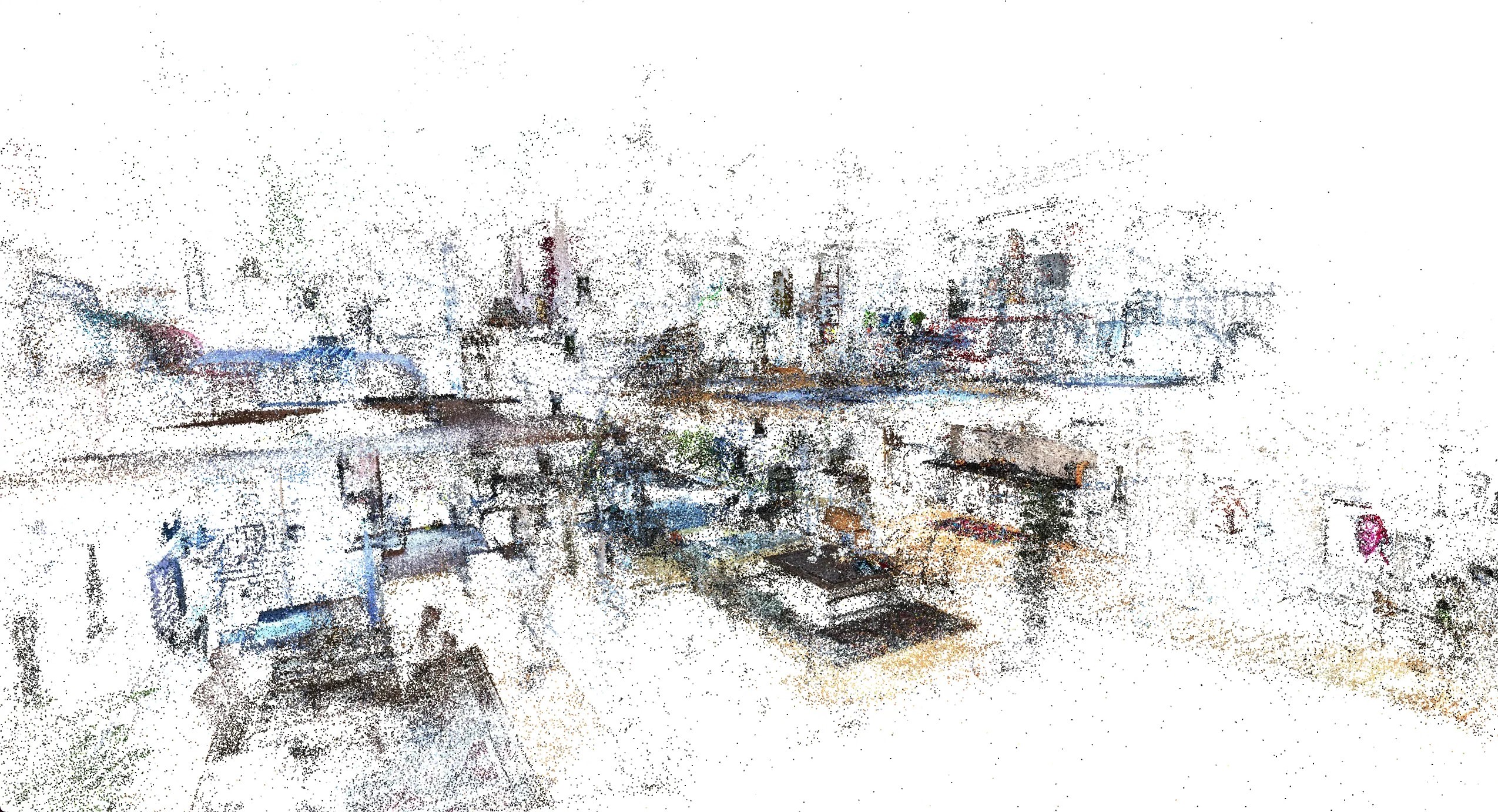}
\caption{A point cloud visualization of map data for a single family home. Different points in the map are associated with different physical spaces and can leak information about the structure of surrounding objects and the environment.}
\label{fig:point_cloud_example}
\end{figure*}

We define a \textit{virtual space} to be all map data associated with a physical space, 
and a \textit{private} space to be a space with certain access restrictions.
The focus of our work is to protect private virtual spaces against unauthorized access and tampering from adversaries.
We assume users will configure access control policies over the spaces they own to protect them from unauthorized access by other users, and that these policies are provided as input to VACMaps.

In the physical world, private physical spaces are often associated with an owner or manager who poses access restrictions on other users.
We expect virtual space to have similar concepts of access control and ownership but without some of the physical limitations.
Unlike in the physical world, in AR users can move 
across virtual spaces and establish \textit{digital presence} as they interact with AR services.
If an unwanted user establishes digital presence in virtual spaces, it raises confidentiality and integrity concerns.
Examples of harm caused by unwanted access to maps via AR systems includes spying (confidentiality concerns) and vandalism (integrity concerns).

\textbf{Confidentiality}
There are two potential concerns with confidentiality in AR maps: accessing raw map data and 
performing unauthorized localization.

First, map data captures information about object structure that can be used to infer what is around the user and general information about the surrounding environment~\cite{su2015, point-cloud-object-classification, hackel17}.
To illustrate this concern more concretely, \autoref{fig:point_cloud_example} shows a point cloud visualization of the map for a canonical single-family home.
In this visualization, an observer can infer details about the house layout~\cite{nguyen2013, wang2018} and different objects~\cite{su2015} using existing computer vision algorithms.
Prior work~\cite{invsfm, dangwal21, mahendran2015, weinzaepfel2011} has shown point clouds are not entirely secure and that point cloud data from a stolen map can be reversed to reproduce a surprisingly good visual reconstructions of the environment.
If accessed by an unauthorized adversary, the map data can leak confidential information about what objects a user has or is in the surrounding environment.

Second, the ability to localize into an unauthorized virtual space using a map raises additional confidentiality concerns.
Once the adversary has localized to a private space, they have access to a rich rendering of the setting.
This could, for example, enable passive spying on objects in the space, e.g., a computer screen, without being physical present similar to existing spyware. 
Strong read access control over maps can rule out both of these cases.

\textbf{Vandalism and Integrity}
Unauthorized modification of map data for private virtual spaces is a 
concern for similar reasons: tampering, adding, or deleting AR objects compromises a virtual space's integrity.
When mapping or updating a physical space, it is necessary to prevent unauthorized adversaries from degrading integrity by deleting or modifying it.
Prior work has shown that degrading the integrity of point cloud models or maps can negatively impact the accuracy of downstream tasks or inject backdoor triggers~\cite{sun2021, point-cloud-backdoor-attack, wicker2019}.
Such attacks are loosely the virtual analog of an adversary stealing objects (deleting map data) from or posting/placing items (tampering/adding map data) in one's physical space, respectively.
Deleting or spamming map data directly degrades the quality of many AR services that rely on map integrity.

Adversaries may also attempt to modify or delete the map data of a victim. 
For instance, a shop may attempt to override or modify the map of a competitor to gain a commercial advantage by degrading map integrity.
In a personal setting, without proper map write access control, AR users would be vulnerable to adversaries spontaneously adding objectionable content in their spaces, which may not be easy to delete without proper permissions.


\section{Threat Model and System Goals}
\label{subsec:threat-model}

This section defines the system assumptions, adversary model, and objectives of VACMaps.

\subsection{System Definition}

We define our system as having two primary entities: AR device users (clients) and the service provider (server or cloud).
We assume each user in this system owns an ego-centric (i.e., head-mounted) AR device that is responsible for collecting the visual information required for conducting server-based localization and mapping.
These user devices interface with a cloud service provider that updates the virtual map of the world. The 
service provider is assumed to be honest, since it is incentivized to maintain accurate and up-to-date map 
data and localization estimates for users; inaccurate localization estimates or stale map data would 
degrade the quality of AR experiences that rely on the integrity of the map.

The primary data asset in our system is map data, which consists of 3D point locations and their associated feature descriptors. 
In server-based localization and mapping, the virtual map of the world is stored on the service provider as it is impractical to store all of it on any individual device; we assume that existing network security protocols and best practices ensure the channel between the device and the service provider is secure.
During mapping, when a user moves through an unmapped space, the user device will generate map data and upload it to the service provider.
The service provider will then attempt to merge that data into the existing map as long the user has correct permissions to write data for that physical space.
If a user tries to localize into a mapped space, they will request read access to the relevant map data for that physical space based on GPS information.
If the environment has changed when a user moves through this space, they may also provide updates to it which again requires write access.
Depending on the access control policies defined over the physical spaces a user is in, users may or may not be able to map, update, or access this map data.

We assume that there is map data of shared spaces that a user may want to share with others as well as map data the user would like to restrict access to (e.g., their house or apartment).
In our system architecture, users cannot directly share maps with each other and all map updates and localization requests go through the server.
This also means that users cannot store entire maps on their local devices.
We also assume that users may want to access maps to localize in spaces that another user may have already mapped.
This allows users to amortize the cost of mapping the environment as a shared map allows users to align their view of the world with a crowd-sourced version to enable AR use cases.

\subsection{Adversary Model}

We assume that a subset of AR device users will behave maliciously and either attempt to perform unauthorized map accesses to either read private data or modify/delete points of other users\footnote{We assume that the malicious user is from the public and is not mentioned in any allow policy for the map, therefore does not have permission to access the maps under any condition.}.
A malicious adversary is able to manipulate the visual information and GPS data that is captured on the AR device before it is uploaded to the service provider.
Falsifying GPS data enables an adversary to potentially trick the localization and mapping system into serving maps to localize into areas a user may not physically be present.
Falsifying visual information allows an adversary to directly fabricate feature points and hence map data that can be combined with GPS spoofing to attempt to write invalid map data.
However, an adversary need to model precisely the visual information as if the adversary is physically present in an environment to trick the mapping and localization system, which comes at a high cost.
Therefore, we assume that an adversary is able to
spoof GPS data but not visual information.
Finally, we assume that malicious devices are not able to further compromise other devices in the systems, including the datacenter service provider, and that the adversary is not able to fake their identity.

\subsection{System Objectives}

The key objective of our system is to protect user map data of their private spaces from unauthorized access by adversaries; we define unauthorized access as any read, modification, or write request that is prohibited by a set of access control policies (see \autoref{sec:architecture}).
We assume that there are policies defined over physical spaces that govern the mapping, sharing, and modification of map data over these spaces which the system must comply with.
AR is still an emerging technology where the concrete access control policies are still the subject of debate so precisely defining the ``correct'' policy over map data is beyond the scope of this paper.
Rather, we will assume that there are \textit{some} policies that exist which define how spaces are mapped, read, and modified and that they will potentially change over time\footnote{We assume that the spaces (corresponding to rooms in the physical world) are fixed in this work and the policy remains unchanged when processing a single access request.}. 
This means that the access control system should be sufficiently flexible to support and enforce a range of reasonably practical policies over map data.

The objective of the system is to enforce access control over map data such that each access is compliant with these access control policies.
Our system should allow access requests to map data by users if it is permitted under the defined access control policies.
However, it should deny access for any user who may intend to read, modify, or map spaces that are prohibited by the access policies.
For instance, if a user restricts all other users' access to the map of their apartment, an adversary should not be able to read or modify that portion of the map.


\section{VACMaps System Architecture and Language Definition}
\label{sec:architecture}

This section introduces VACMaps and describes how it enforces access control over map data.
We provide a primer on formal methods and a high-level overview of the VACMaps system.
We then present the domain-specific language \policydslname{} for access policies and formalize the semantics of \policydslname{} using SMT formulas,
which enables automated reasoning about policies.
Additionally, we show how VACMaps leverages natural spatial hierarchy in maps to improve performance.

\begin{figure*}[htbp]
  \centering
  \includegraphics[width=0.8\textwidth]{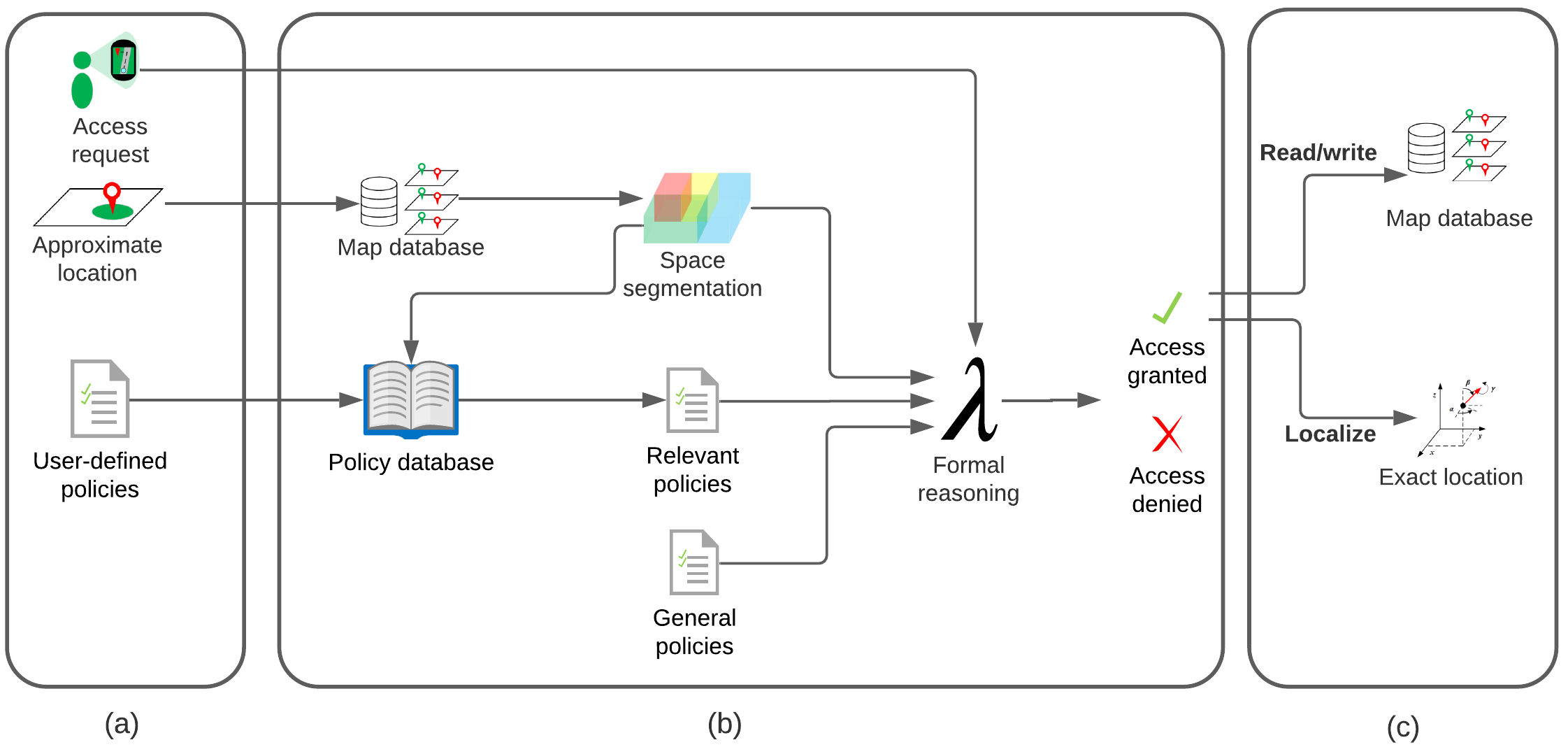}
  \caption{Architecture of VACMaps. (a) A user's device provides a set of user access control policies, approximate location determined by GPS, and access request type (i.e, read or modify). (b) The VACMaps system checks the map database to extract relevant maps and the 3D segmentation of the space, as well as relevant policies which are processed by formal reasoning to determine whether access should be denied or granted. (c) If access is granted for the request, the user request is serviced appropriately.}
    \label{fig:system-architecture}
\end{figure*}

\subsection{Formal Methods Primer}

Formal methods study techniques for \textit{specification} and \textit{verification}~\cite{clarke1996formal}. 
Specification involves defining a system or process and its desired properties using a mathematically defined syntax and semantics.
Verification on the other hand is a technique that proves
that the system or process satisfies certain properties.
Formal methods are often used to prove the correctness of safety-critical systems,
e.g., a flight collision avoidance system~\cite{heimdahl1996completeness} or
systems that are much too complicated to verify by hand (e.g, a modern compiler
toolchain~\cite{appel2011verified,kastner2018compcert}).

Formal method starts by representing a system (e.g., a computer program)
as formulas in Boolean logic and defines a Boolean satisfiability problem that
encodes the property we wish to verify (e.g., reachability, liveness, or
correctness). The Boolean satisfiability solver (SAT solver) then tries to
decide whether the Boolean constraints combined with logical connectives can be
made true by choosing true or false values for each variable. 
Depending on whether the answer is true or
false, this information can be used to make claims about the system. 

Boolean logic
itself is often not expressive enough to encode problem that requires domain
knowledge. As a result, the Boolean SAT problem is often augmented to involve
predicates on integers, reals, and data structure in addition to Booleans using
satifiability modulo theory (SMT)~\cite{de2011satisfiability}. As an example,
one might ask whether there exist integers $x, y, z$ such that
$x > 2y \wedge y > z \wedge \neg (2z <= x)$ is true. The meaning of predicates like $x > 2y$,
i.e., given symbols $x, y$ decide if $x > 2y$ should be true or false, is
provided by the \textit{theory of linear integer arithmetic} which defines the semantics of integer math. For the previous query, the solver would verify that the formula is unsatisfiable for any values of $x$, $y$, and $z$. 
Other theories include theory of uninterpreted functions, theory of strings, etc. 
In this work, we focus on formulas within theory of linear real arithmetic with equality for which there exists efficient decision procedures.

The underlying SAT problem itself is a well-known NP-Complete problem; however,
modern solvers implement sophisticated heuristics that exploit the power of
modern processors and the structure of formulas to explore the search space very
efficiently. As a result, modern SMT solvers like CVC~\cite{cvc4}, Yices~\cite{yices}, and 
Z3~\cite{z3} can often handle formulas with hundreds of thousands of variables in a
reasonable amount of time~\cite{de2011satisfiability}.

This also means SMT solvers are fast enough to support practical sized problems in areas such as program
verification and testing~\cite{farzan2015automated}, compiler optimization
correctness~\cite{kundu2009proving}, interactive theorem proving~\cite{bertot2013interactive}, and
program synthesis~\cite{srivastava2010program}. Problems in specific application domains are
translated into SMT formulas such that the satisfiability of the formulas
implies that certain properties must hold. For example, we can over-approximate the
the states that a program could run into, i.e., maintain a set of states that is
guaranteed to be a super set of the actual reachable states.
Then we might encode the event that a particular error state is within this set
of reachable states as an SMT formula so that if the formula is unsatisfiable then the program is
guaranteed to \textit{never} run into that error; and if the formula is
satisfiable we could obtain from the SMT solver a concrete example of an initial
configuration of the program that \textit{may} run into that error (since we
maintain an over-approximation of reachable states rather than the exact set of reachable states).

In our work, we describe the semantics of access policies formally using SMT formulas.
Properties of the access policy semantics imply the correctness of the access
control system given any combination of access policies and access requests.
We will also encode certain claims about access policies as queries to SMT solvers.
By examining the SMT solver output, we can then detect possible policies misconfigurations
and report it to the users.

\subsection{VACMaps Architecture Overview}
\label{subsec:architecture-overview}

\autoref{fig:system-architecture} depicts the VACMaps system.
VACMaps provides access control for localization and mapping in the client-server setting.
We assume that users have AR devices with cameras that capture visual observations of the physical world that are converted to map points. 
We further assume the user has defined access policies over the physical spaces they own/control and the virtual maps they create. 
Together, the observed map points and access policies serve as the input to VACMaps.

On the server side, localization and mapping services compute the user's exact current
location and map point coordinates using input from the user's device.
Notice that GPS, WiFi, and other measurements that provides the approximation location are not directly used to decide the user's location but only as an aid to the localization and mapping services.
The server also checks if the user's observations match the database to ensure that the user is not spoofing its location. 
Additionally, the server maintains a database of access policies and the mapping between policies and spaces\footnote{We assume that in practice there will be a separate system that manages the ownership of spaces which is able to decide who is allowed to modify existing policies}.
These policies are either 
generic\footnote{Default policy configurations that are not user-specific (e.g., deny access to all private spaces by default).} or user-defined. 
The policy database extracts the list of relevant policies needed to determine the user's access rights.
These policies are organized in a spatial hierarchy to make identifying relevant policies efficient (see \autoref{subsec:spatial-hierarchy}).
VACMaps takes the relevant access policies, and 
the calculated exact location of the user to formulates a SMT problem.
The problem is finally presented to the formal verification engine to verify whether the map access request
should be granted or denied (see \autoref{subsec:formal-reasoning-policies}).

Map access requests are modeled as read, write, and localize in VACMaps.
If the user has read access, a virtual map is visible. 
If the user has write access, the 3D feature points that the user
observed can be used to update the known map or create (write) a new one, which is stored on the server.
If the user has localize access, the AR device could show the
user's relative location with respect to the known map.

\begin{figure}[ht]
  \centering
  \footnotesize
\begin{mdframed}
  \setlength{\grammarparsep}{5pt plus 1pt minus 1pt} 
\begin{grammar}
  <policy> ::= (<name>, <effect>, <principal>?, <action>?, <space>, <condition>?)

  <name> ::= `Name:' <string>

  <effect> ::= `allow' | `deny'

  <principal> ::= `Principal:' <string>

  <action> ::= `read' | `write' | `localize'

  <space> ::= `Space:' <space-expr>

  <space-expr> ::= <space-id> | $\neg$ <space-expr> | <space-expr> $\wedge$ <space-expr> | <space-expr> $\vee$ <space-expr>

  <condition> ::= `Condition:' <cond-expr>

  <cond-expr> ::= <atom> | $\neg$ <cond-expr> | <cond-expr> $\wedge$ <cond-expr> | <cond-expr> $\vee$ <cond-expr>

  <atom> ::= `TODAfter:' <time> | `TODBefore:' <time> | `WhenInside' <space-id>
\end{grammar}
\end{mdframed}
\caption{\label{fig:policy-dsl-syntax} Syntax for the policy language. Here
  a policy is an ordered tuple of fields, and ``?'' denotes an optional field.
  If an optional field is missing, then the policy applies universally regardless of the value of that field.
  For example, a policy without a ``principal'' statement applies to all principals.
}
\end{figure}

\subsection{\policydslname{}: A Domain-Specific Language for Access Policies}
\label{subsec:dsl-policy}

VACMaps develops a domain-specific language (DSL), named \policydslname{}, for access control policies that defines access rights over virtual maps. 
\autoref{fig:policy-dsl-syntax} shows the abstract syntax for the VACMaps policy DSL. 

In \policydslname{}, a \texttt{policy} contains one or more \texttt{statement}s. 
Each \texttt{statement} has a \texttt{name}, which we use to refer to the policy when changing or deleting policies.
\texttt{Statements} have \texttt{effects} that declare whether this policy is granting or denying access. 
A \texttt{statement} also contains a \texttt{principal} $P$ that indicates the name or ID of the user that the policy applies to.
The \texttt{principal} in a statement is optional; 
omitting the \texttt{principal} allows one to specify a policy that applies to \textit{all} users.
For instance, we might want to deny all access requests to private spaces
such as restrooms (line 20-25 in \autoref{fig:example_policy}).
The \texttt{action} field lists the access events granted or denied by this policy and can be one of three types: 
\texttt{read} -- request map points for the space; 
\texttt{write} -- update the map of the space; 
\texttt{localize} -- determine the user's location in this space.

\texttt{Space} specifies the set of spaces that a policy applies to.
\policydslname{} allows policy designers to specify access to spaces as 
(Boolean) logical expressions.
For example, one can say a policy grants access to ``the first floor of the house
except the bathroom; and the living room on the second floor'' or ``third floor
of the office building and the lobby''.
In VACMaps, we allow \texttt{user} ($U$) access to some \texttt{space} ($S$) if and only if there exists an allowing \texttt{policy} ($P$) that grants $U$ access to $S$ \textit{and} that there is no denying policy that prevents $P$ from accessing $S$.
In other words, deny policies will always override allow policies in VACMaps if there is a conflict.
We assume a space has deny access rights by default.

The \texttt{condition} field is an optional qualifier that defines conditions under which the policy takes effect. 
We typically use conditions to determine access based on a user's current location and time of the day. 
For example, you might allow your friends to map your house only when they are in the house; 
similarly, a museum might only allow visitors to localize into its exhibition spaces only during business hours.

\subsection{Access Control and Spatial Hierarchy}
\label{subsec:spatial-hierarchy}

It is natural for us to think about access rights to objects \textit{hierarchically}. For example, when declaring a folder as read-only one
expects the files inside also become read-only. 
Similarly, users might want to set certain policies on their entire house and expect the rooms inside  to automatically ``inherit'' some of the policies. 
Defining policies in a way that is coherent with respect to the
natural hierarchy of spaces also relieves users' burden to define unique policies
for each individual space they control. 
Organizing policies according to the hierarchy of spaces makes the VACMaps system more efficient.
For example, fetching relevant policies from the policy database based
on the user's current location would be easier to implement given
that we store the policies according to the hierarchy of spaces.

We formalize the idea of an access control system that respects spatial
hierarchy as follows. We assume that if space $a$ and space $b$ has a non-empty
intersection, then either $a$ is contained in $b$ or $b$ is contained in $a$.
We then construct a graph $G = (V, E)$ such that the node set $V$ corresponds to the set
of spaces, and an edge $(u, v) \in E$ if and only if space $v$ is contained in space
$u$. Since any space $u$ can have at most one parent, $G$ is a forest. We say an access
control system is said to be coherent with respect to spatial hierarchy $G$ if a
policy defined for space $S$ is also enforced by the system on space $T$ if $T$
is a descendant of $S$ in forest $G$. VACMaps is a system that is coherent with
respect to spatial hierarchies.

\autoref{fig:room_hierarchy} gives a visualization of the hierarchy of rooms of a canonical single family home.
For example,
the policies set on ``house'' will be forced on all floors and all rooms that
are descendants of the ``house'' node in the graph; policies set on ``Master suite'' will also be enforced on ``Master bedroom'' and ``Master bathroom''
since these spaces are contained inside ``Master suite''. 
One could imagine that
the management of an office building could implement a similar hierarchy for
different floors that belong to different companies that have offices in the
building.

Having a hierarchical structure of spaces also makes access auditing process
more intuitive for users. Users and policy experts can look at the hierarchy
diagram and make queries like ``who has access to this particular space'' and
``visualize spaces that this user access'' and compare the result against their
access control expectations and/or social norms. We will introduce more features of
VACMaps that facilitates policy auditing in
\autoref{subsec:formal-reasoning-policies}.

\subsection{Processing Access Requests and Reasoning about Policies}
\label{subsec:formal-reasoning-policies}

We now describe how VACMaps handles access requests and caches them to improve performance.
The server maintains a mapping from spacial identifiers to policy names that, given a space
$S$, returns all allow and deny policies that refer to $S$ in the \texttt{space-expr}. 
Given a new access request against map point $p$, VACMaps first finds all spaces $S_1 \subseteq S_2 \subseteq ... \subseteq S_n$ in the spatial hierarchy that contain the map point $p$.
VACMaps then translates all applicable policies to spaces $S_1, \dots, S_n$ into an SMT formula $F$ and combines them. 
Policies are translated to SMT formulas in a lazy manner, i.e., they are
only translated into SMT formulas when there is an incoming access request that
involves space (or subspace) to which this policy may apply. 
The access request is then evaluated against $F$ resulting in an allow or deny decision.
To improve systems efficiency, VACMaps also maintains an LRU cache for access permissions.
The data for the cache is a tuple consisting of identifier for the smallest enclosing space $S_1$ for point $p$,
as well as current time of day and other contextual information in the access request (with the exception
of the map point coordinates) and the access decision.
Once VACMaps processed access request for point $p$, other access requests against map points in space $S_1$ with the same contextual information
will be served using the cache.

\begin{figure*}[htb]
    \centering
    \begin{subfigure}[ht]{0.3\textwidth}
         \centering
\includegraphics[width=\linewidth]{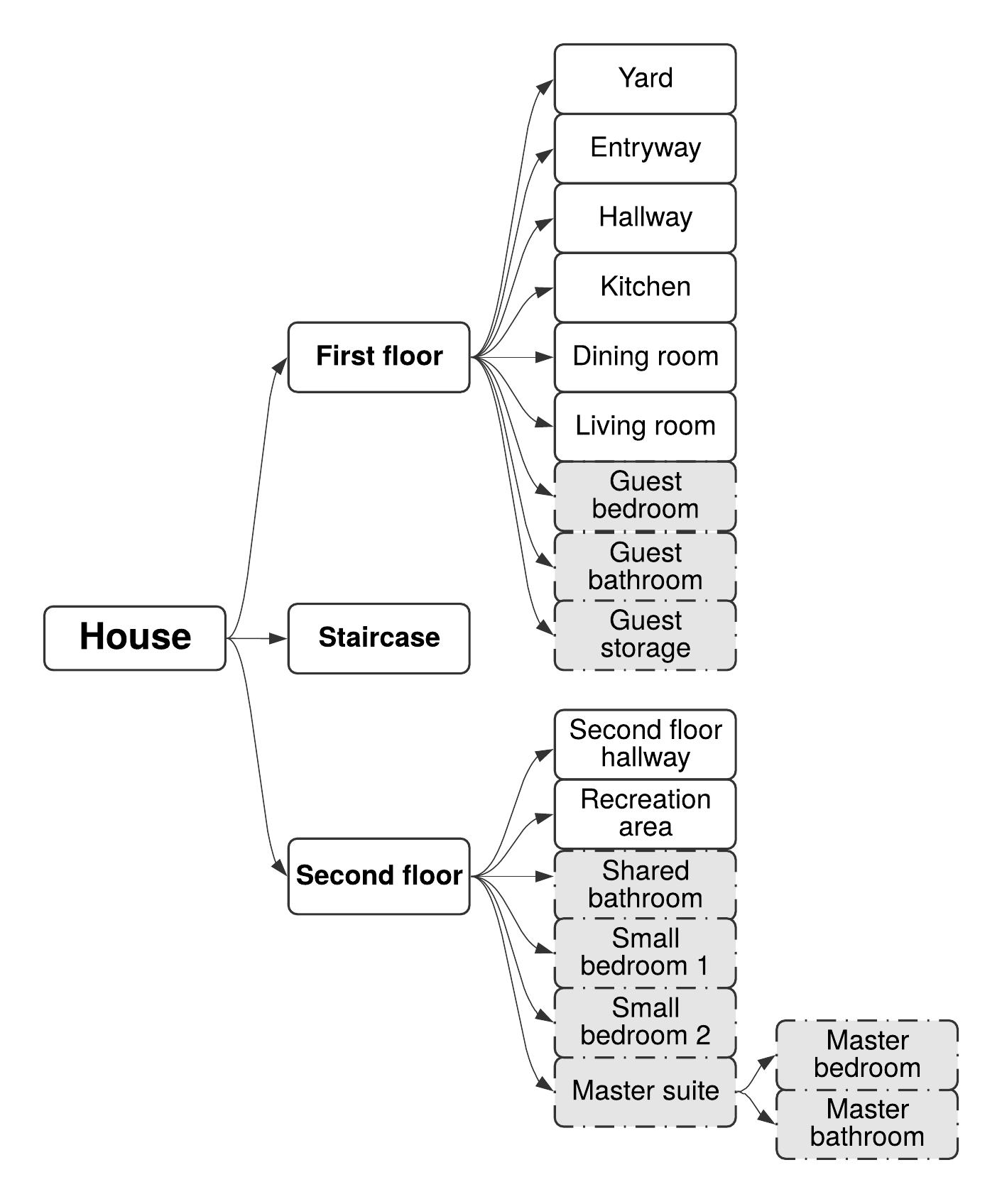}
\caption{
Hierarchy of rooms and spaces in our single-family
     home dataset. Shaded boxes correspond to rooms/spaces for which we want to
     implement certain access-control mechanisms that prevent certain residents of
     the house from mapping these rooms.
\label{fig:room_hierarchy} }
       \end{subfigure}
       \hfill
     \begin{subfigure}[ht]{0.3\textwidth}
         \centering
 \begin{lstlisting}
Begin
Name: "GrantAliceAllAccess"
Effect: allow
Principal: "Alice"
Action: read
Space:    "first_floor_all" Or "second_floor_all"
       Or "staircase"
End

Begin
Name: "GrantBobAccessToGuestArea"
Effect: allow
Principal: "Bob"
Action: read
Space: recreation_area Or small_bedroom_2
Condition: UserInside: "second_floor_all"
           And TODAfter: 0900
End

Begin
Name: "DenyAccessToBathroom"
Effect: deny
Space:    "guest_bathroom" Or "shared_bathroom"
       Or "master_bathroom"
End
\end{lstlisting}
  \caption{An example set of access control policies for the single-family house dataset. The first policy grants Alice read access to all spaces
    in the house. The second policy gives Bob read access to the recreation area and a small bedroom
    on the second floor given that Bob is in the house and the current time is after 9 am. The third policy
    prevents anyone from accessing the bathrooms.
    \label{fig:example_policy} }
     \end{subfigure}
           \hfill
 \begin{subfigure}[ht]{0.3\textwidth}
         \centering
         \tiny
         \begin{align*}
           f_{\text{first policy}} \triangleq & \quad s_{\text{principal}}= \text{``Alice''} \\
                                       &\wedge s_{\text{action}}= \text{``read''} \\
                                       & \wedge ((x, y, z) \in B_{\text{first floor}} \\
                                       & \quad \vee (x,y,z) \in B_{\text{second floor}}) \\
           f_{\text{second policy}} \triangleq 
                                   & (s_{\text{principal}}= \text{``Bob''} \\
                                       &\quad \wedge s_{\text{action}}= \text{``read''} \\
                                       &\quad \wedge ( (x,y,z) \in B_{\text{small bedroom 2}} \\
                                       &\quad \quad \vee (x,y,z) \in B_{\text{recreation area}} ) \\
                                       &\quad \wedge (u_{x}, u_{y}, u_{z}) \in B_{\text{second floor}} \\
                                       &\quad \wedge  0900 \leq t \leq 2400) \\
           f_{\text{third policy}} \triangleq &\quad \neg( ((x, y, z) \in B_{\text{guest bathroom}} \\
                                       &\quad  \vee (x, y, z) \in B_{\text{shared bathroom}} \\
                                       &\quad \vee (x, y, z) \in B_{\text{master bathroom}}))
         \end{align*}
         \caption{ Example SMT formula translation for the policies over three
           rooms. Notice that each room inherits the policy of
           its enclosing space. Also, Alice still does not have access to the
           bathroom since her access rights are ``overridden'' by the deny
           policy on the bathroom that applies universally to all principals.
           For a query point on the second floor, its relevant policies are 
           represented as an SMT formula $(f_{\text{first policy}} \vee f_{\text{second policy}}) \wedge f_{\text{third policy}} $.}
         \label{fig:smt_translation_of_eg_policy}
       \end{subfigure}
       \caption{Spatial hierarchy, access policy, and SMT formula translation of policy on the
       example house dataset. Here the master bedroom is the rightmost bedroom on the second
           floor of the house, small bedroom 2 is one of the small bedrooms on
           the second floor, and guest bathroom is the bathroom on the first
           floor. For readability we use an abbreviation $(x, y, z) \in B$ to
           represent the conjunctive formula $l_{x} \leq x \leq r_{x} \wedge \dots$
           specifying 3D point with coordinate $(x, y, z)$ must lie within an
           axis-aligned box $B$. }
      \label{fig:hierarchy_policy_translation}
\end{figure*}

\autoref{fig:smt_translation_of_eg_policy} shows how VACMaps translates \policydslname{} policies to an SMT formula.
First, we discuss the translation for an allow policy. 
The formula translation of an allow policy is done by building a
conjunction of the translation of each of its statements, except for the effect statement. 
To translate the principal statement, we declare a string-typed
variable $s_{\text{principal}}$ and we translate \texttt{principal: ``Alice''} into
an equality between the string-typed variable and a string literal:
$s_{\text{principal}} = \text{``Alice''}$. 
The action statement is translated similarly.
Next, we translate the space statement. 
We declare real-typed variables $x, y, z$ representing the coordinate of an incoming map point. 
To translate the atomic expression, which consists of simply a space identifier, we first look up
the boundaries of the space. VACMaps assumes all spaces are axis-aligned
bounding boxes, so the boundaries can be represented as a tuple
$\langle l_{x}, r_{x}, l_{y}, r_{y}, l_{z}, r_{z}\rangle$. The atomic expression then
becomes $l_{x} \leq x \leq r_{x} \wedge l_{y} \leq y \leq r_{y} \wedge l_{z} \leq z \leq r_{z}$. This can
naturally be lifted into a translation of an arbitrary space statement. For the
condition statement, we also only need to specify how we translate the atoms.
The \texttt{WhenInside} atom could be translated in a way similar to space
expressions provided that we define real-typed variables $u_{x}, u_{y}, u_{z}$
representing the user's approximate location where the access request is generated. We
translate the \texttt{TODBefore} and \texttt{TODAfter} atoms by introducing a
integer-typed variable $t$ and translate an condition that encodes ``after 9 pm
and before 1 am'' into formula $(2100 \leq t \leq 2400) \vee (0000 \leq t \leq 0100)$. To
translate a deny policy, we simply take the negation of the conjunction of the
translation for its statements.
We give an example of a policy in \autoref{fig:example_policy} and its SMT
formula encoding in \autoref{fig:smt_translation_of_eg_policy}.

An access request $\langle v_{\text{principal}}, v_{\text{action}}, x, y, z, u_{x}, u_{y},\allowbreak u_{z}, t \rangle$
is a tuple that contains the name of the principal making the request, the action
the principal wants to perform, the coordinate of a map point $(p_{x}, p_{y}, p_{z})$ in 3D space,
the location of the user $(l_{x}, l_{y}, l_{z})$, and current time of day $t$.
To evaluate an access request, we find all spaces that contain the map point
being queried and obtain a formula $F$ encoding the combined policy for all
these spaces. We then perform a substitution in $F[s_{\text{principal}} \mapsto v_{\text{principal}},
s_{\text{action}} \mapsto v_{\text{action}}, (x, y, z)\mapsto (p_{x}, p_{y}, p_{z}), (u_{x}, u_{y}, u_{z}) \mapsto (l_{x}, l_{y}, l_{z}) ]$
and then perform a bottom-up simplification of the resulting formula.
The result can only be \texttt{true} or \texttt{false} since every atom
is evaluated to either \texttt{true} or \texttt{false} and the resulting
formula are these Boolean values connected via logical connectives $\wedge, \vee, \neg$.
We will introduce other analyses we do on the SMT representation of
the policies in \autoref{sec:implementation-and-eval}.

\section{Methodology and Evaluation}
\label{sec:implementation-and-eval}

This section outlines the implementation methodology and evaluates VACMaps's flexibility, scalability, reliability and performance.

\subsection{Methodology}

We implement VACMaps in Python using \texttt{textX} to implement and parse our
access control policy DSL.
We use \texttt{intervaltree} to define and maintain the spatial hierarchy relationship of spaces; we also use Z3 (\texttt{z3-solver}) as our solver backend.
All experiments were run
on a MacBook Pro 2019 with a $2.4$ GHz $8$-Core Intel Core i9 CPU and $32$ GB RAM.

\begin{figure*}[htb]
    \centering
    \begin{subfigure}[t]{0.3\textwidth}
         \centering
         \includegraphics[width=\textwidth]{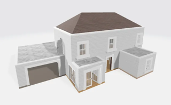}
         \caption{Outside look of the house.}
         \label{fig:house_outside}
       \end{subfigure}
       \hfill
     \begin{subfigure}[t]{0.3\textwidth}
         \centering
         \includegraphics[width=\textwidth]{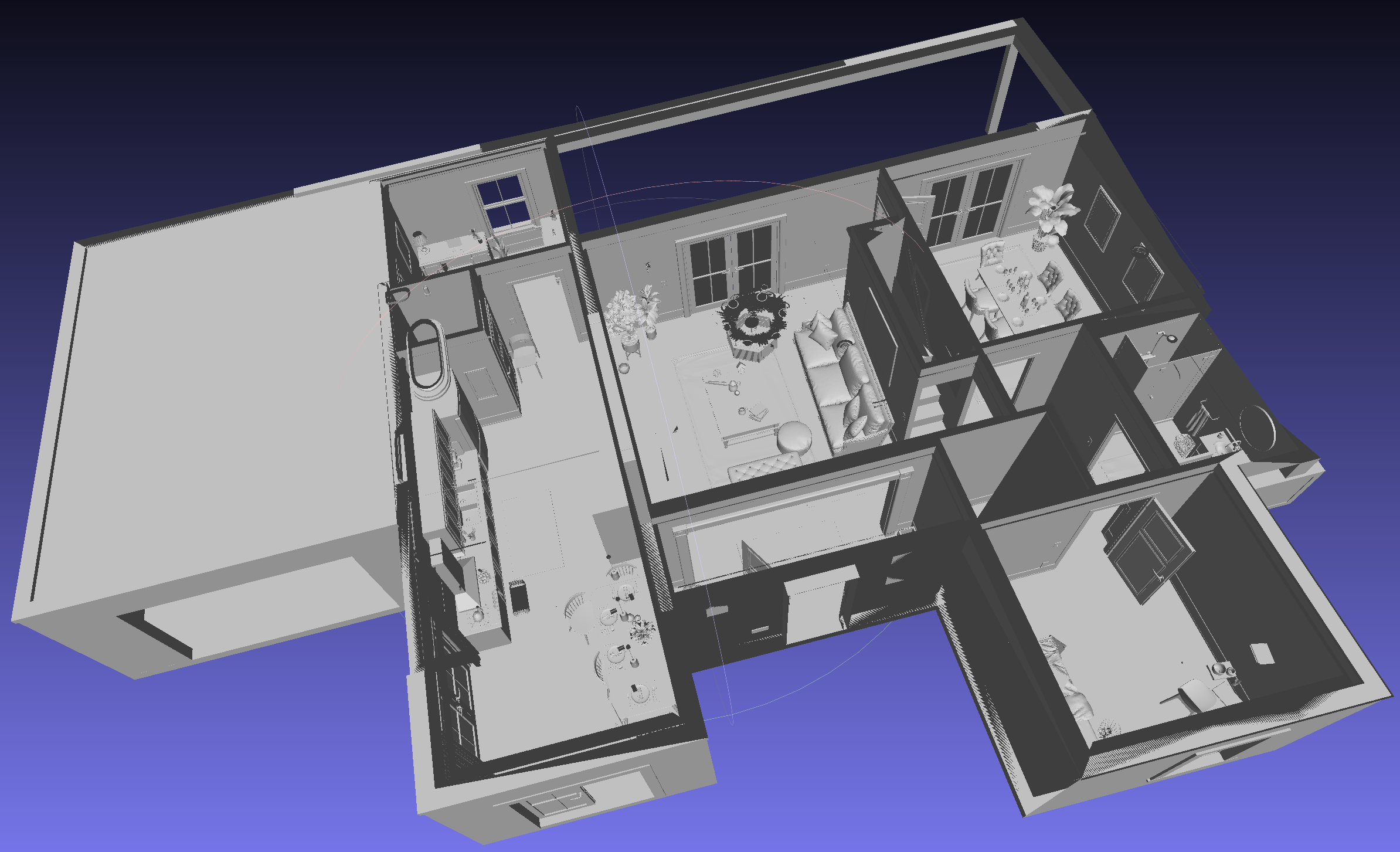}
         \caption{Floor plan for rooms on the first floor.}
         \label{fig:house_first_floor_rooms}
       \end{subfigure}
       \hfill
     \begin{subfigure}[t]{0.3\textwidth}
         \centering
         \includegraphics[width=\textwidth]{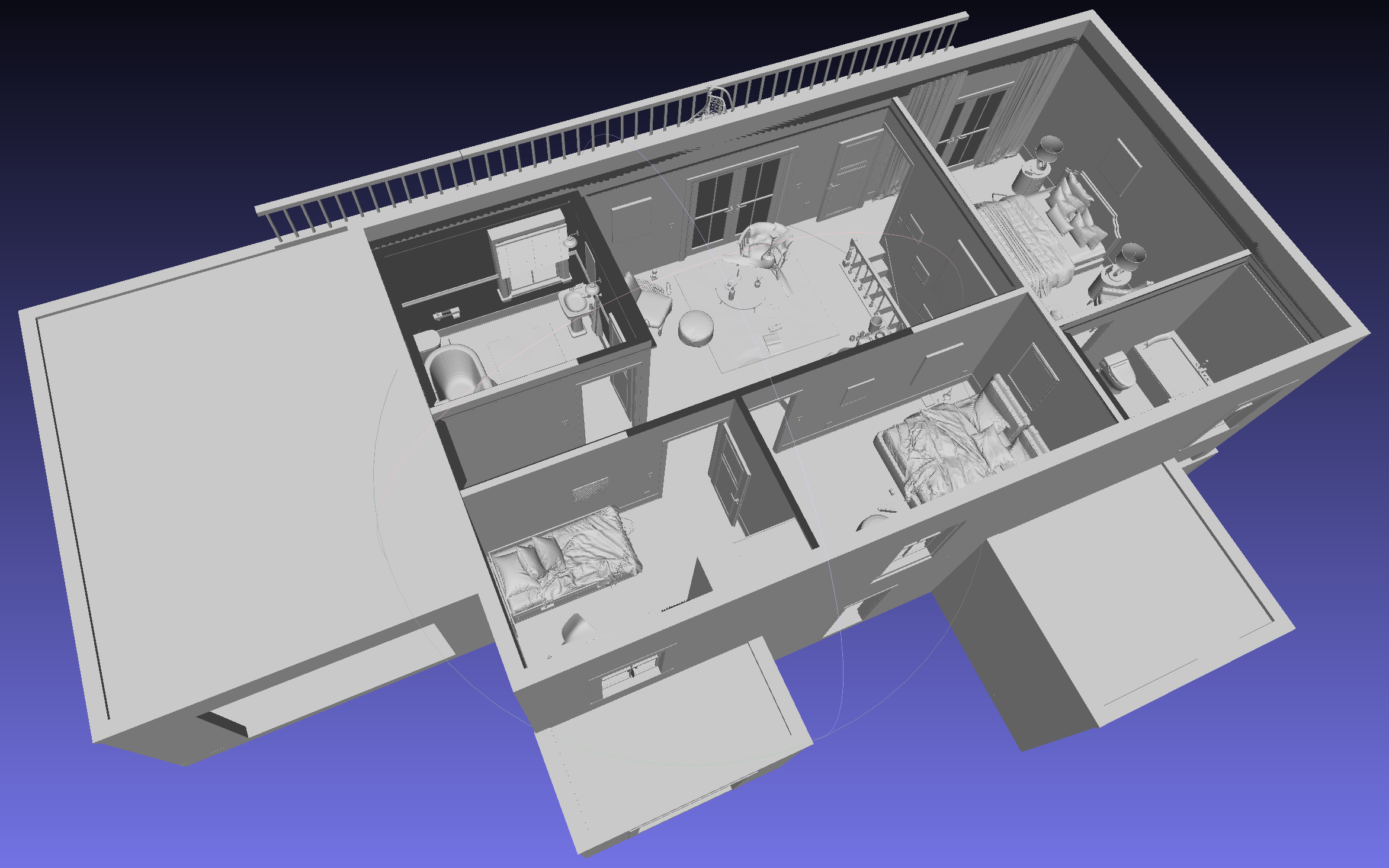}
         \caption{Floor plan for rooms on the second floor.}
         \label{fig:house_second_floor_rooms}
     \end{subfigure}\\[1ex]
     \begin{subfigure}[t]{0.3\textwidth}
         \centering
         \includegraphics[width=\textwidth]{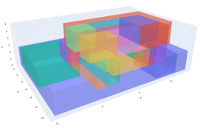}
         \caption{3D room segmentation of the house where each 3D bounding box corresponds to a room.}
         \label{fig:room_seg}
     \end{subfigure}
     \hfill
     \begin{subfigure}[t]{0.3\textwidth}
         \centering
         \includegraphics[width=\textwidth]{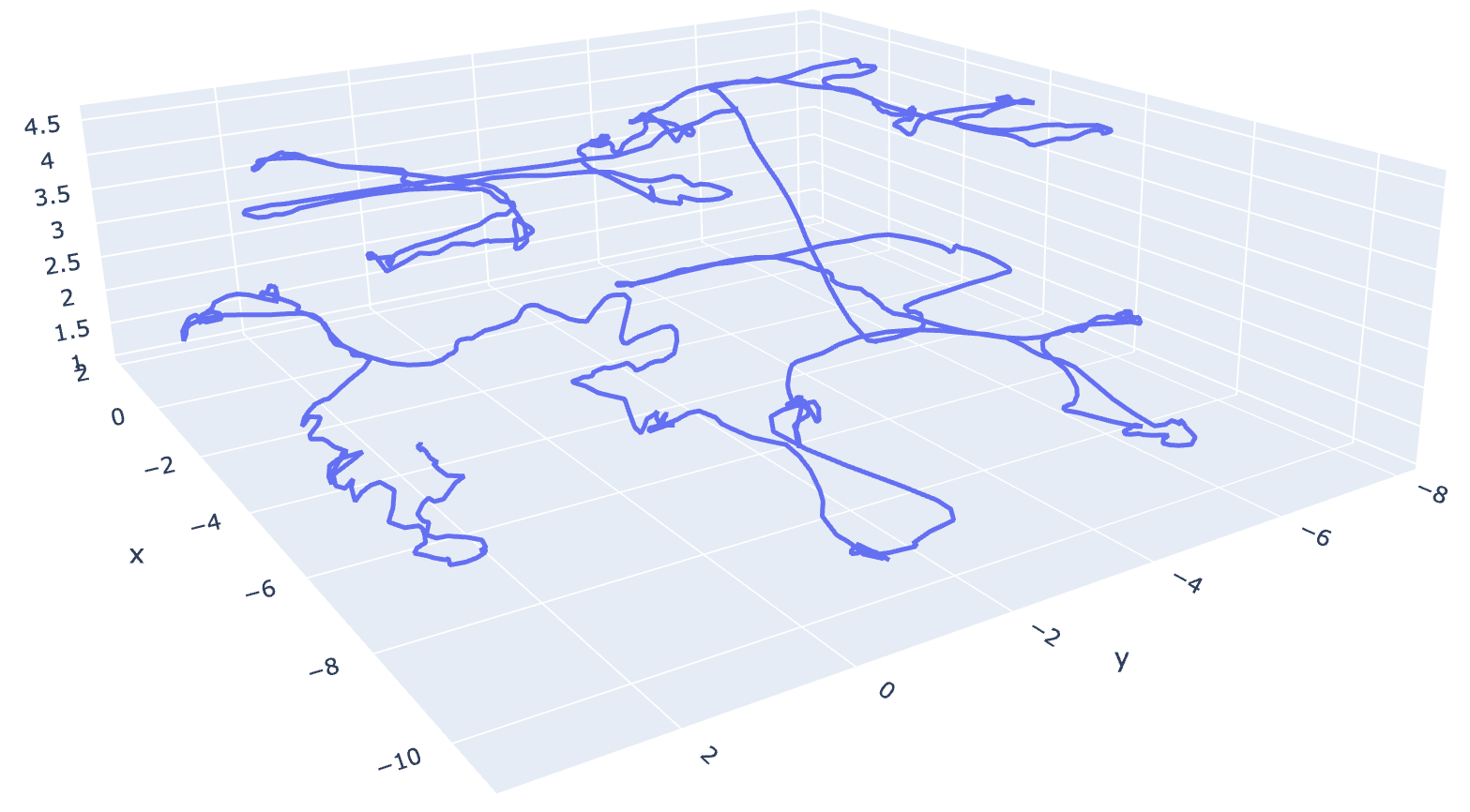}
         \caption{A simulated tour through the house which is used to generate access sequences to map points. The tour starts on the
           first floor, visits every room on the first floor, and
           then visits every room on the second floor.}
         \label{fig:house_tour}
    \end{subfigure}
     \hfill
     \begin{subfigure}[t]{0.3\textwidth}
         \centering
         \includegraphics[width=\textwidth]{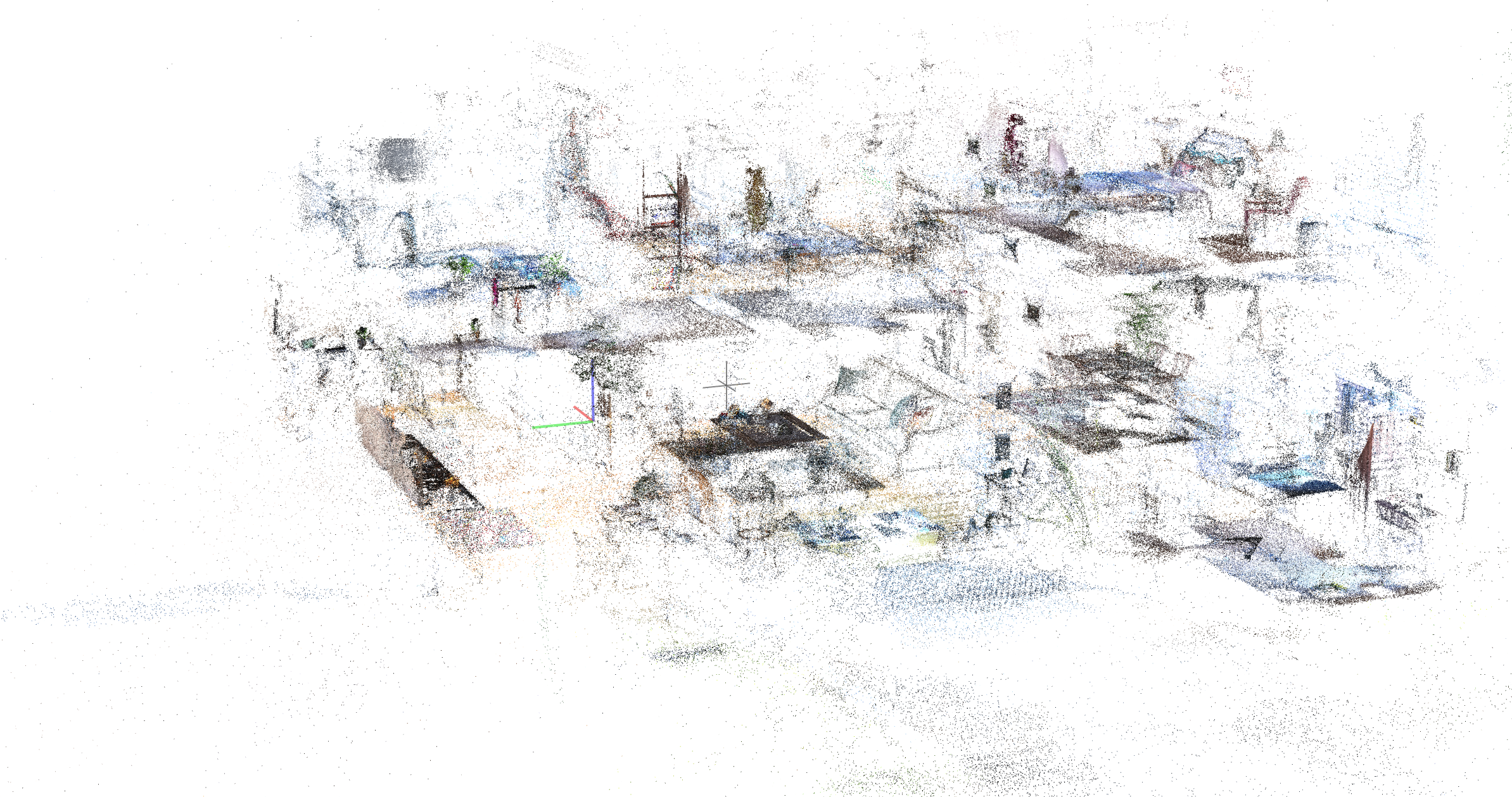}
         \caption{A point cloud representing feature points observed and captured during the simulated tour. We want to decide the user's access rights to these points based on the policies defined for the house.
         }
         \label{fig:house_point_cloud}
     \end{subfigure}
     \caption{Visualization of the single-family house dataset used to evaluate VACMaps. The dataset provides ground truth (b-c) floor plans, (d) 3D space segmentation, (e) a simulated tour, and (f) a map of observed points observed during the tour.}
    \label{fig:sfh_dataset_visualization}
\end{figure*}

To our knowledge, there is no standardized benchmark for evaluating access
policies for localization and mapping or general AR/VR applications. 
We thus use
a synthetically generated dataset for a single family house
(\autoref{fig:sfh_dataset_visualization}) that captures common
scenarios that a user would encounter when they interact with an AR device
at home.

The house has two stories and contains a set of rooms shown in \autoref{fig:room_hierarchy}.
We obtain the 3D segmentation and the name of the rooms
(\autoref{fig:room_seg}) using the ground truth floor plans used to construct
the house. In practice, this can be obtained using techniques like PlaneRCNN~\cite{liu2019planercnn}.
The dataset also provides a video sequence of a tour or walk through (\autoref{fig:house_tour}) of all the rooms in the house to simulate
what the camera attached to the user's AR device would capture.
This effectively generates a sequence of accesses to the map which VACMaps validates with respect to access policies.

The localization and mapping pipeline takes in as input the simulated video
sequence. For each frame in the video sequence, the pipeline returns a list of
feature points (both their 2D coordinates in the image and their 3D coordinates
in space) that are used to decide the user's location or update the map
database. The pipeline also returns the user's estimated location for each
frame.

We assume that users send \textit{access requests} to the access control system
to request access rights to map points for each frame of the simulated video. An
access request simply contains the user's ID, all the feature points that the user
observes at this time instant, the user's current location, and other metadata
the AR device generates (such as the timestamp) that are useful in deciding
access rights.

Access policies for rooms in the house are hand-written (unless otherwise specified) to verify the
correctness and evaluate the performance of VACMaps. We will
discuss how we create the policies in the following subsections.

\subsection{Basic Access Control Scenario Evaluation}

We first evaluate the basic utility and functionality of VACMaps for canonical
use case scenarios. 
Each of these basic scenarios that a user may encounter is briefly described in
\autoref{tab:scenarios}. 
We implement all these scenarios by crafting a set of access control policies over the rooms and using a subset of the frames in the tour that reflect the test scenario.
We then inspect
that the access request decision (i.e., allow/deny) that VACMaps makes is
consistent with the expected result. 
We use these scenarios to validate the
basic functionality of VACMaps as an access control system.


\begin{table*}[htb]
\centering
\caption{Description of common access control scenarios used to verify VACMap's basic functionality and correctness. \label{tab:scenarios} }
\begin{tabular}{@{}p{0.25\linewidth}p{0.35\linewidth}p{0.35\linewidth}@{}}
  Scenario & Description & Expected Access Behavior \\ \midrule
  Conservative default policy & A space does not have any defined access policy associated with it. & All users are by default denied access to this space. \\
  Private spaces & User A is inside their own private space. & Only user A should be allowed access to this space. \\
  Shared private spaces & A user is going to the restroom. & User should be denied access for mapping and localization. \\
  Bystander spaces & User A is walking by user B's house where the windows are open and A can see into B's private space. & User A should be denied the ability to map user B's private space. \\
  Localize with another user's map & User A tries to localize himself into an environment using map points generated by user B. User B wants to keep their map data private though. & User A should be denied access to user B's map. \\
  Denying friends of friends access & User A invites a friend user B who invites user C who is a friend of B but not a friend of A. User A trusts user B but not user C with mapping the house. & User B should be able to map and localize but user C should be denied access. \\
  Private space map contamination & User A uses a compromised device to inject map points into user B's private space. & Access requests by user A to modify user B's map should be denied. \\
  Changing or revoking access policies & User A wants to change the condition under which a friend B can map A's house, and wants to revoke user C's access to map A's house. & Once the policy change request is handled, user B and C's access rights to A's house are modified and enforced through the access control system. \\
\end{tabular}
\end{table*}

\begin{figure*}[htb]
  \centering
    \begin{subfigure}[t]{0.15\textwidth}
         \centering
\includegraphics[width=\textwidth]{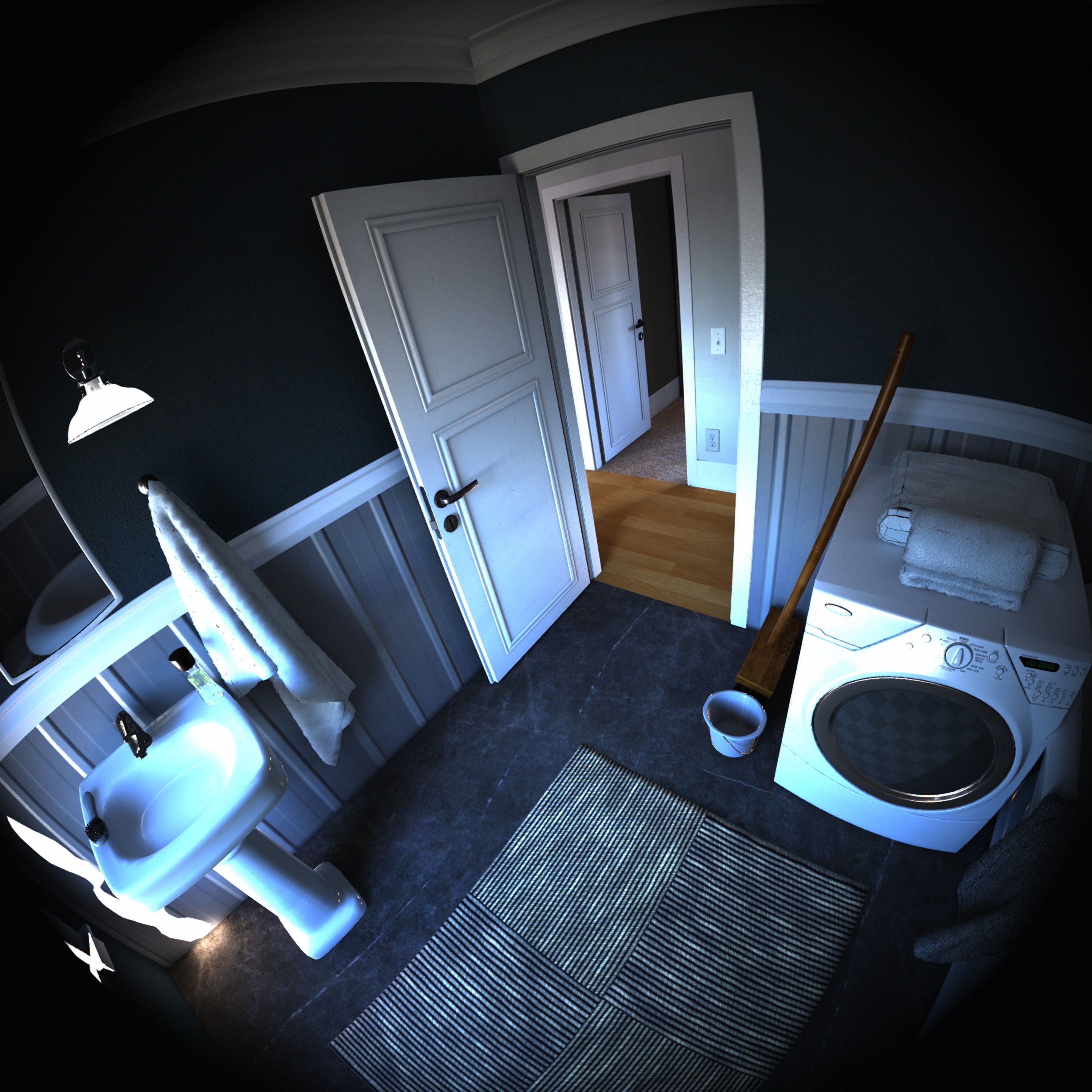}
\caption{A picture taken in the second floor bathroom.   \label{fig:bathroom-pic} }
\end{subfigure}
       \hfill
     \begin{subfigure}[t]{0.3\textwidth}
         \centering
         \includegraphics[width=\textwidth]{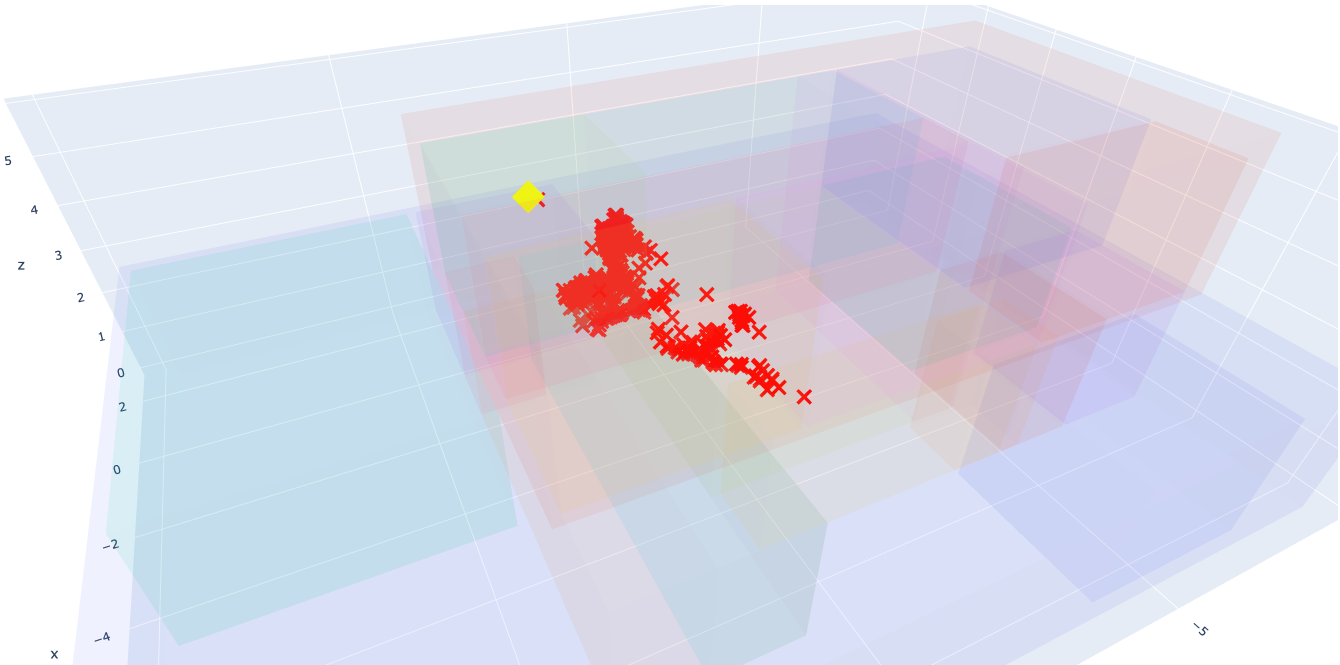}
         \caption{Shared Private Spaces: VACMaps denies access to map points localized to be
           in the bathroom.}
         \label{fig:bathroom-acc}
       \end{subfigure}
       \hfill
 \begin{subfigure}[t]{0.15\textwidth}
         \centering
\includegraphics[width=\textwidth]{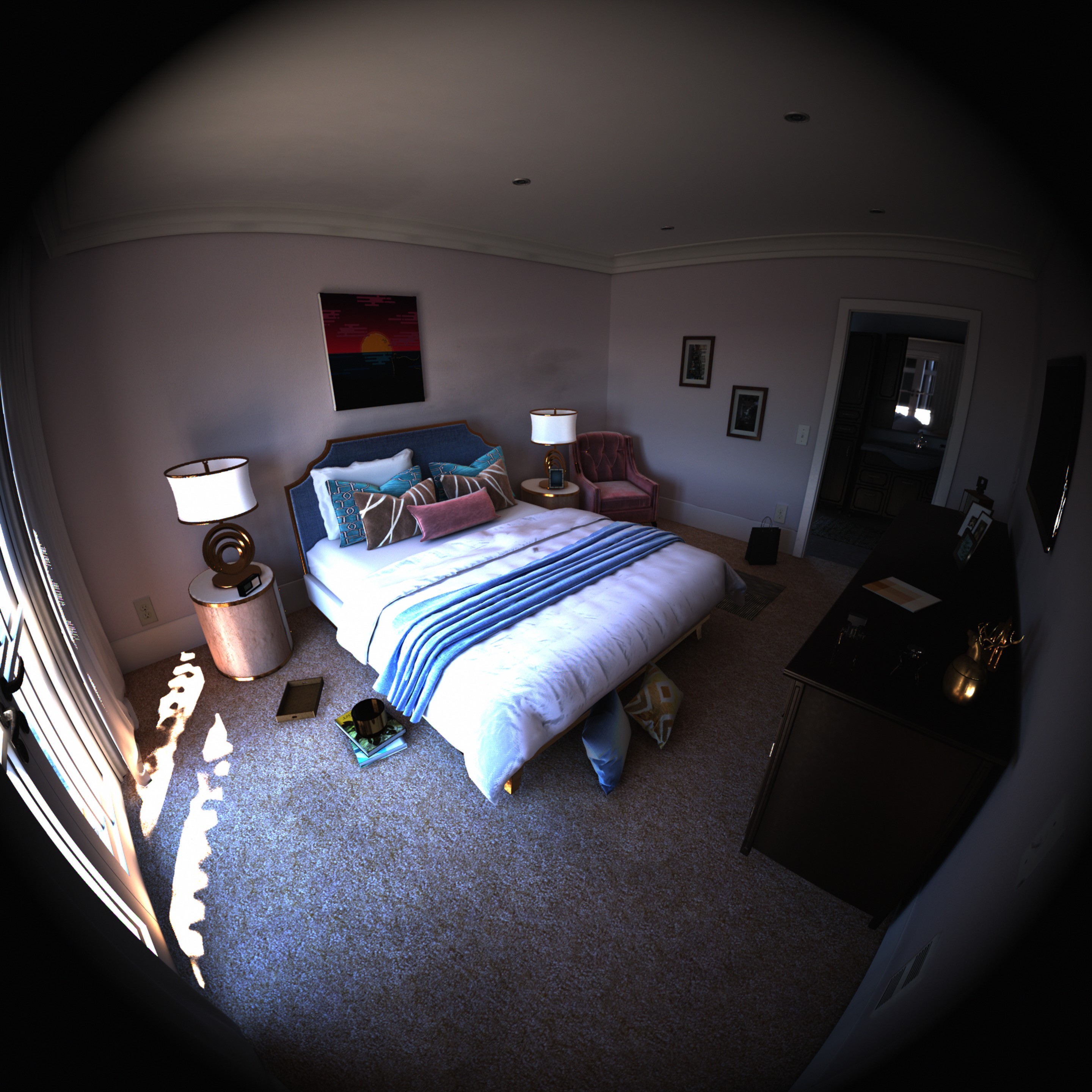}
\caption{A picture taken in the second floor master bedroom.
    \label{fig:private-bedroom-pic} }
       \end{subfigure}
       \hfill
     \begin{subfigure}[t]{0.3\textwidth}
         \centering
         \includegraphics[width=\textwidth]{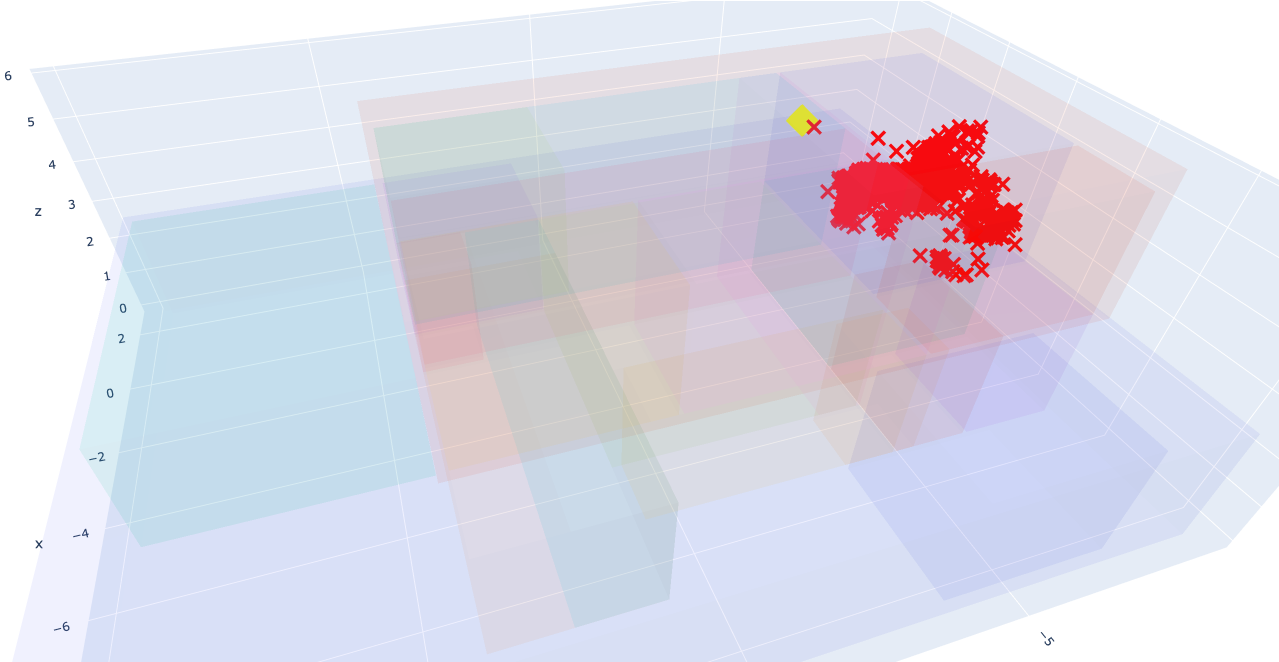}
         \caption{Private Spaces: VACMaps denies access to map points for a different user's bedroom.}
         \label{fig:private-bedroom-acc}
       \end{subfigure}
       \\ [1ex]
 \begin{subfigure}[t]{0.12\textwidth}
         \centering
\includegraphics[width=\textwidth]{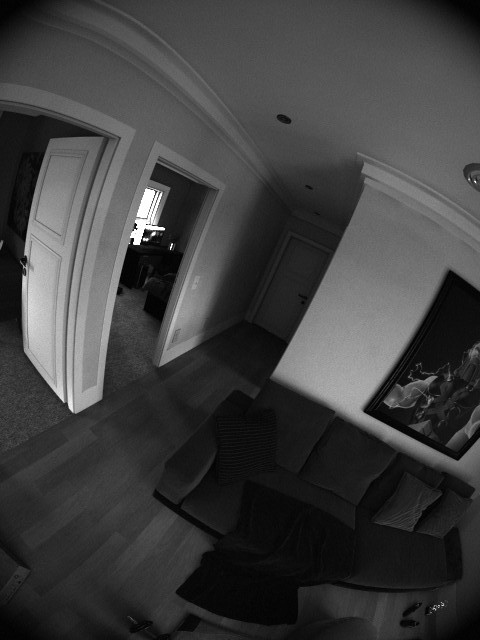}
\caption{A picture taken in the second floor common area \label{fig:mixed-pic} }
       \end{subfigure}
       \hfill
     \begin{subfigure}[t]{0.32\textwidth}
         \centering
         \includegraphics[width=\textwidth]{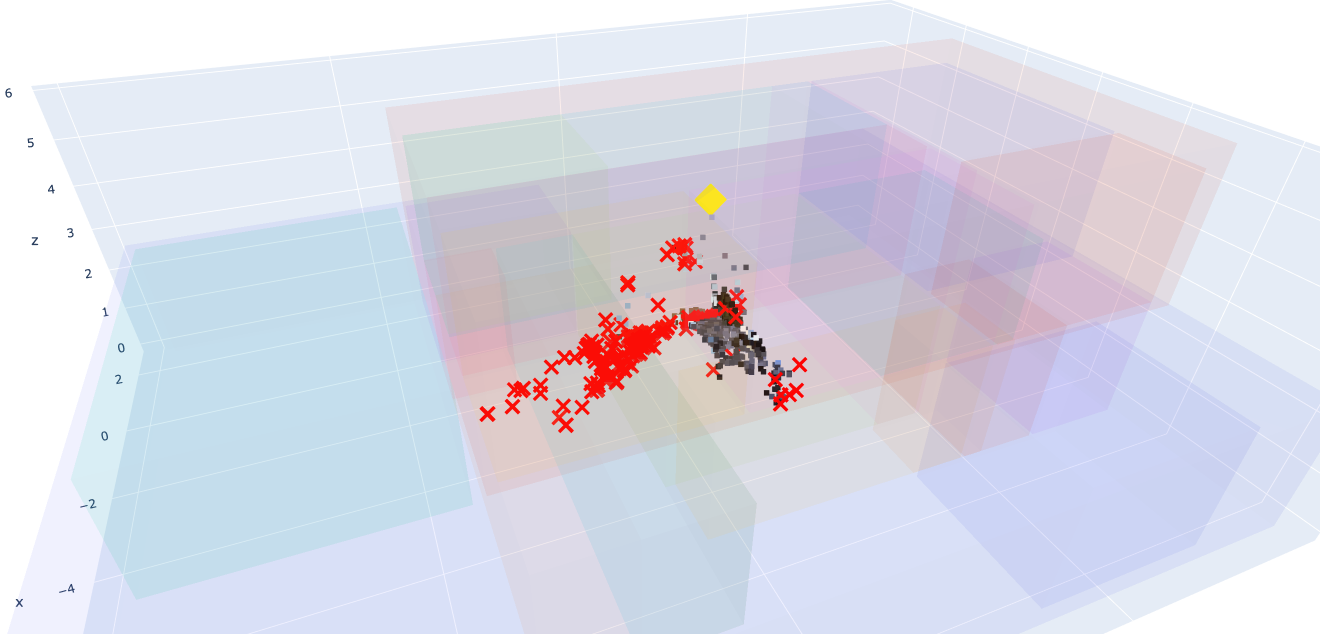}
         \caption{Bystander Spaces: VACMaps only allows user access to points that lie in the common area
           but not in another user's bedroom.}
         \label{fig:mixed-acc}
       \end{subfigure}
       \hfill
 \begin{subfigure}[t]{0.12\textwidth}
         \centering
\includegraphics[width=\textwidth]{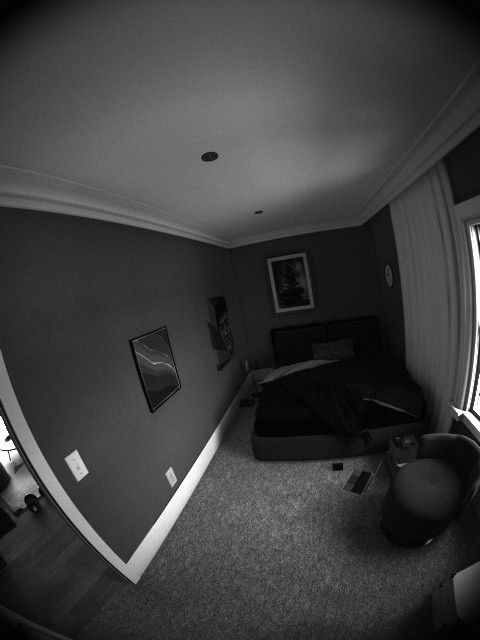}
\caption{A picture taken in the user's bedroom.
  \label{fig:own-bedroom-pic} }
       \end{subfigure}
       \hfill
     \begin{subfigure}[t]{0.32\textwidth}
         \centering
         \includegraphics[width=\textwidth]{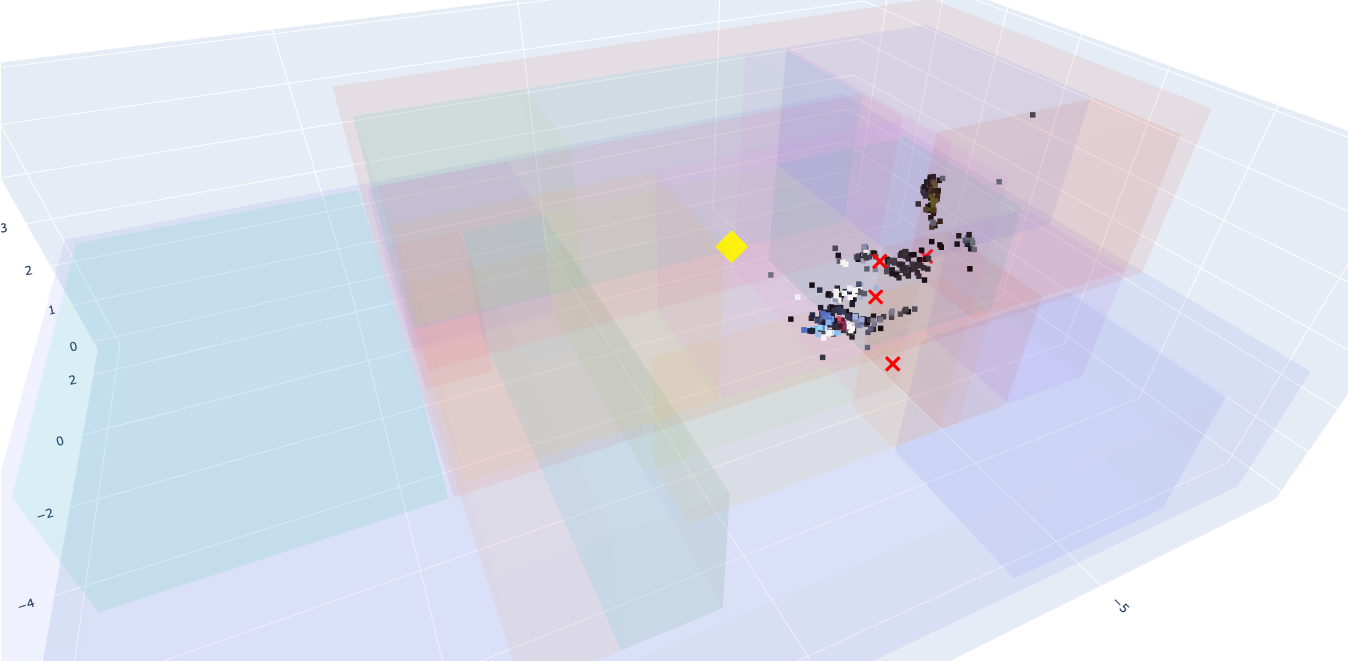}
         \caption{Private Spaces: VACMaps allows access to points inside the user's own bedroom.}
         \label{fig:own-bedroom-acc}
       \end{subfigure}
       \caption{Visualization of access rights determined by VACMaps. Map points are marked with
       its RGB value if access is granted and are displayed as a red cross otherwise. The yellow diamond denotes where the AR device is located.}
    \label{fig:visualization_access_rights}
\end{figure*}

The results for several of these scenarios is shown in \autoref{fig:visualization_access_rights} where RGB points denote points that the user is granted access to while red Xs denote points to which a user is denied access.
\autoref{fig:bathroom-acc} illustrates the the scenario of shared private spaces where a user is looking into the bathroom.
When a user tries to access these map points VACMaps denies access to all map points in the bathroom as indicated by the red map point annotations in the diagram.
In \autoref{fig:private-bedroom-acc}, the user is located in a different user's private bedroom; similar to the previous bathroom scenario, VACMaps again denies access to these map points because the user is not authorized to view or use them.
On the other hand, if the user is located in their own private space VACMaps appropriately allows the user access to the map (\autoref{fig:own-bedroom-acc}).
Finally, \autoref{fig:mixed-acc} illustrates the case where there is a bystander space which may be private but is next to a common area space which is not.
In this case, both the bystander space and common space are within the user's field of view so a user may request access to both of them.
VACMaps appropriately denies access to the points located in the private bystander spaces while allowing access to map points in the common area.

\subsection{Performance and Scalability}

\begin{figure*}[htbp]
    \centering
    \begin{subfigure}[t]{0.48\textwidth}
         \centering
\includegraphics[width=\linewidth]{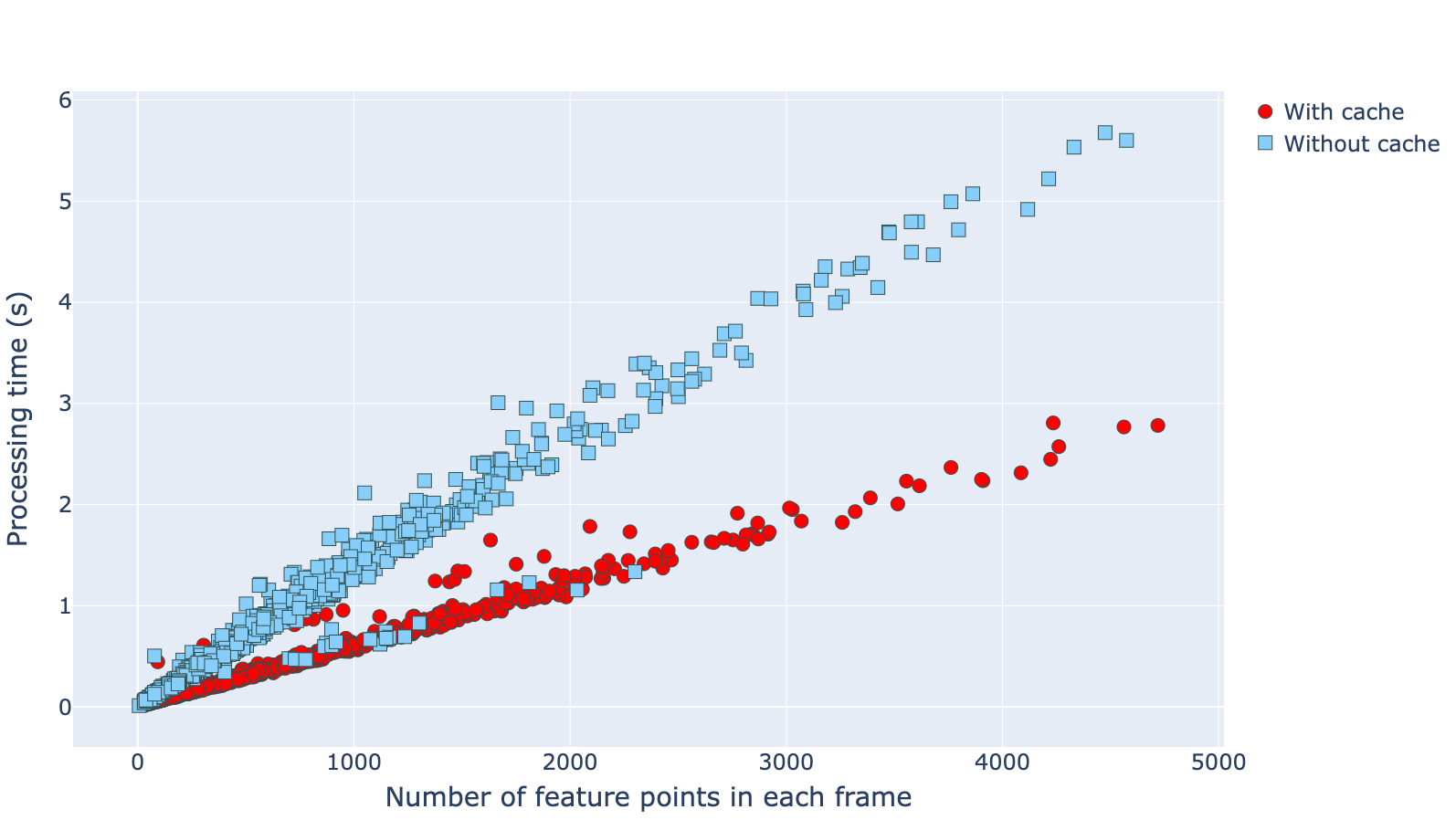}
  \caption{Run time of VACMaps to process one frame as the number of
    extracted feature points increases in the
    image, with (red dots) and without (blue dots) the LRU cache implementation.\label{fig:running_time_vs_num_points} }
       \end{subfigure}
       \hfill
     \begin{subfigure}[t]{0.48\textwidth}
         \centering
         \includegraphics[width=\textwidth]{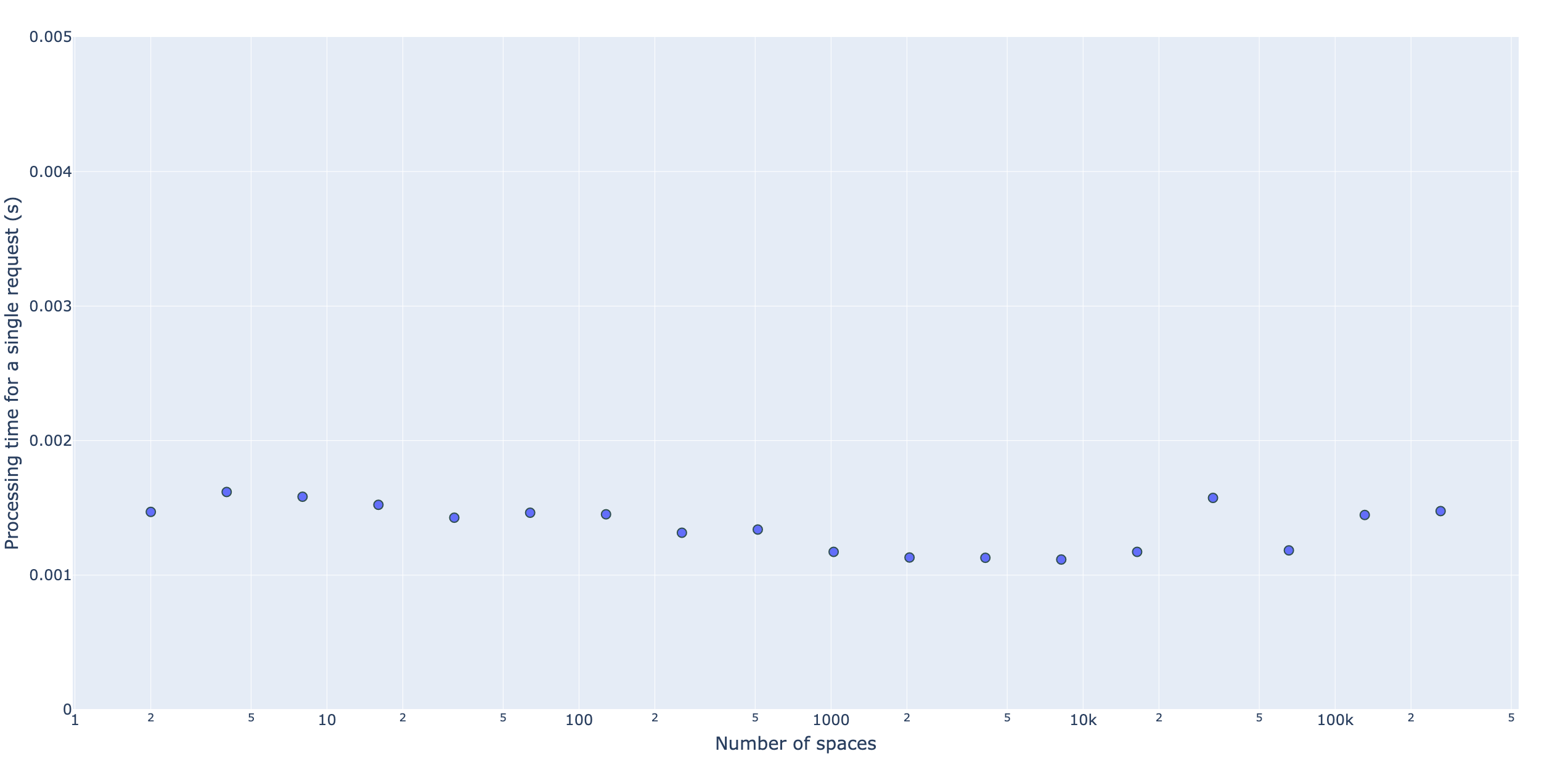}
         \caption{The spatial hierarchy implementation in VACMaps scales to handle a large number of spaces:
         processing time does not change much even for $100$k spaces.}
         \label{fig:scalability_num_spaces}
       \end{subfigure} \\ [1ex]
     \begin{subfigure}[t]{0.45\textwidth}
         \centering
         \includegraphics[width=\textwidth]{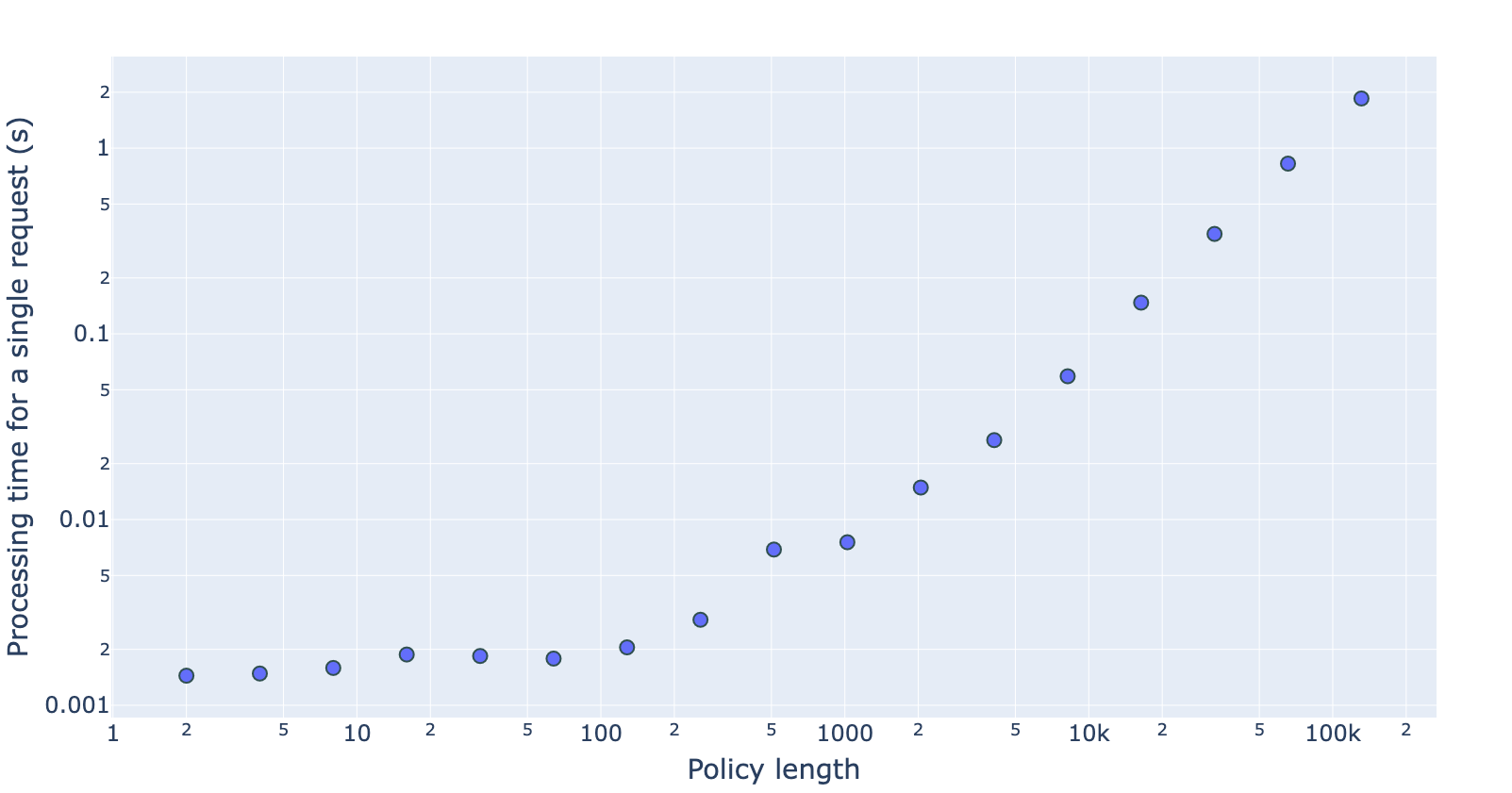}
         \caption{VACMaps scales to handle complex policies. Processing time grows linearly when policy length is large enough. It is likely that the reason for the smaller slope in the beginning is that solvers implement optimizations for short formulas that are not possible for long formulas } 
         \label{fig:scalability_length_policies}
     \end{subfigure}
     \hfill
\begin{subfigure}[t]{0.45\textwidth}
         \centering
         \includegraphics[width=\textwidth]{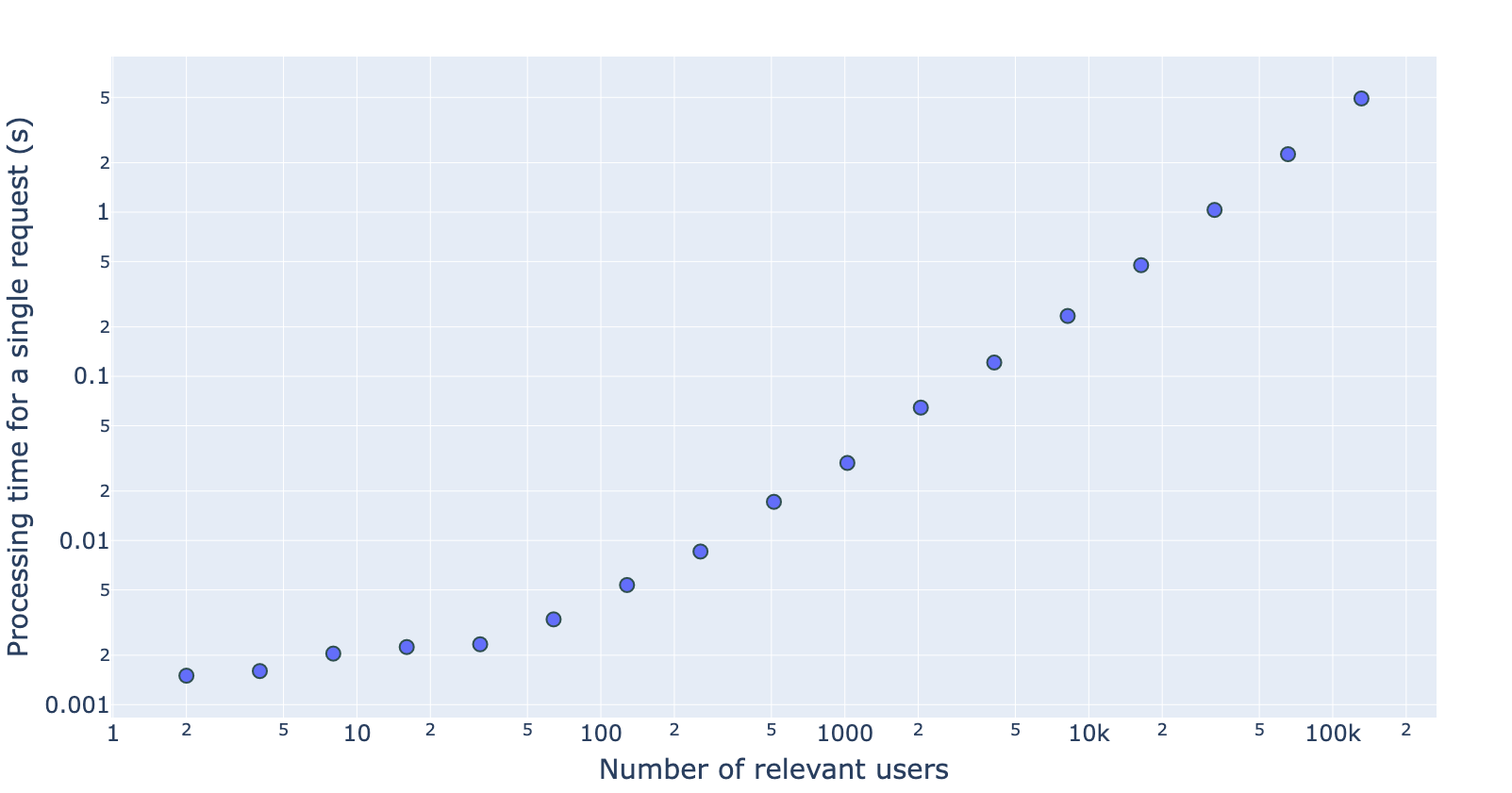}
         \caption{VACMaps scales linearly to handle a large number of relevant users. Processing time is linear w.r.t. number of relevant users, as expected.}
         \label{fig:scalability_num_users}
     \end{subfigure}

     \caption{
     VACMaps performance and scalability with (a) more features per frame, (b) more virtual spaces, (c) longer policies, and (d) more relevant users.}

    \label{fig:eval_performance_scalability}
\end{figure*}

We now evaluate VACMaps's performance and show how it scales to increasing numbers of policies and virtual spaces.

\subsubsection{Can VACMaps Be Fast Enough in Practice?}

Ideally, access control should not add much
overhead to the run time of localization and mapping. In this experiment,
we evaluate the time needed for VACMaps to handle access requests.
We use the synthetic house dataset (\autoref{fig:sfh_dataset_visualization}) as the
testing environment. We define a series of access policies that allows a user
access to some of the rooms while denying the user access to others. We then
follow the simulated tour (\autoref{fig:house_tour}) and send access requests
for all extracted feature points within a frame as the user to VACMaps. We then
measure the time it takes for VACMaps to process all access request originating
from each frame.

The result is shown in \autoref{fig:running_time_vs_num_points}. As the number
of extracted feature points increases, the total processing time for each frame
increases linearly. Since access requests can essentially be handled in a
parallel manner, the optimized  processing time can be much shorter
in practice.
Since we expect to see many points within the same space in a single frame, we
also implement an LRU cache that stores the results of recent access rights queries.
The performance gain of the cache denote by red dots in \autoref{fig:running_time_vs_num_points} shows the cache roughly halves the processing time.

\subsubsection{Does VACMaps Scale to Handle Large Maps and Complex Policies?}

Localization and mapping must be a service that can scale to potentially
thousands of physical spaces with arbitrarily complex access
policies. To evaluate the scalability of VACMaps, we evaluate the performance of
VACMaps for larger maps with many physical spaces, spaces associated with lengthy
access control policies, and spaces whose access policy involves a lot of users.

First, we study the run time of VACMaps as the total number of spaces $n$ in the
system increases (\autoref{fig:scalability_num_spaces}). We create $n$ unit cubes
as spaces and associate with each cube a simple access policy. Then we measure
the time for VACMaps to serve an access request to a random cube. 
The results in \autoref{fig:scalability_num_spaces} show that the run time does not increase significantly even if we have more than $100,000$ spaces.
We attribute this directly to an efficient implementation of the spatial hierarchy organization that VACMaps employs (\autoref{subsec:spatial-hierarchy}) using 3D segment trees; this implementation enables VACMaps to reduce the complexity of looking up relevant spaces for a 3D point to $O(\log n) + m$ where $m$
is the output size, i.e., number of spaces actually containing the point.
As a result, because of the logarithmic complexity, the resulting performance is effectively the same even as the number of space increases.

We also evaluate the performance of VACMaps as the complexity of policies increases. We create $n$ unit cubes as spaces
and associate cube $n$ with a policy of length $O(n)$. This policy is a large
disjunction of clauses that all evaluates to \texttt{false}. The system needs to
evaluate each clause so we expect the processing time to be $O(n)$ for a single
access request against cube $n$. 
The results in \autoref{fig:scalability_length_policies} illustrate that the run time is roughly linear as the number of policies increases.

Finally, we evaluate the performance of VACMaps
as the number of relevant users grow\footnote{The set of relevant users with respect to a space contains all 
principals mentioned in the policies for that space.}. We create $1$ unit cube as the only space under
consideration and associate with it $n$ policies, each mentions a different user.
This in principle has the same effect as increasing the policy length and
we indeed observe a similar trend in \autoref{fig:scalability_num_users}.

Overall, we find that the scalability of VACMaps across number of features per frame, virtual spaces, policy count, and users is reasonable.


\begin{table*}[htbp]
  \caption{
    Examples of auditing functionalities enabled by VACMaps. VACMaps can craft queries to the SMT engine to execute checks for misconfigurations of interest. The results can then be used to help a user audit potential problems in policy configurations.
  }
  \label{tab:auditing}
  \begin{tabular}{@{}p{0.3\linewidth}p{0.65\linewidth}@{}}
    Query & Implementation \\ \midrule
    List all users who have access to a space under certain conditions. & Query the solver for a satisfying assignments to the variables and output the
                                                                          principal, add a blocking clause to the policy formula to prevent the same user being selected again. Repeat until the formula becomes unsatisfiable. \\
    \hline
    Are policies for a non-public space too weak so that everyone could access it? & Check if the formula representing a weak policy stating that everyone
                                                                                     can access this space implies the formula encoding the existing
                                                                                     policies for this space. \\
    \hline
    Are policies for a private space too strong so that even the owner could not access it? & Check if the formula encoding the policy for the space
                                                                                              conjuncted with a clause that specifies the principal equals
                                                                                              the owner of the space is satisfiable. \\
    \hline
    Does a new ``allow'' policy really extend access rights to some space? & Check if the original policy formula conjuncted with the new policy formula
                                                                              is weaker than the original formula. The check is non-trivial since
                                                                              deny policies override allow policies regardless of the ordered in which
                                                                              they are added. \\
    \hline
    Is there a user who is allowed access to a space in some policy but denied access to the same space by another policy? & Collect the set $S$ of all principals that appear in the allow policies for a space. Any deny policy must have an SMT encoding with form $\neg Q$. We just check if any $n \in S$ can make the formula $Q \wedge s_{\text{principal}} = n$ satisfiable. \\
    \hline
    Are policies for a space more permissive than its enclosing spaces? & Check if the formula representation for the space's policies is weaker than
                                                                          that for enclosing spaces. \\
  \end{tabular}
\end{table*}


\subsection{Auditing and Verifying Access Permissions}

While VACMaps takes access policies as input to verify and enforce access control requests, it does not reason about whether a user properly specified policies in the first place.
This leaves it entirely up to the user to define intended access policies.
Specifying policies can be challenging as it requires meticulously combining conditions across spaces, users, and time.
While VACMaps cannot infer user intent, it can aid users in properly setting and debugging policies.

We enable auditing of privacy policies over spaces leveraging the logic reasoning and SMT formula semantics that VACMaps is built upon.
At a high level, the auditing of policy configurations of a space is done by constructing (potentially a series of) SMT queries 
against the formula representation of policies.
This feature can detect potential policy misconfigurations in existing policies
or prevent problematic policies from being introduced in the first place.
\autoref{tab:auditing} shows a set of common scenarios that a user may want to audit for their virtual spaces and how we
implement these audits within VACMaps.
For example, a user configuring access policies for their home may want to restrict access 
to their living room to family members and thus want to see who has access after entering the policy.
If access is overly permissive, this audit will reveal which other users have access.
The owner can then use this information to fix policy misconfigurations.
As another example, we consider the effect of pre-existing deny policies. 
Since deny policies override allow policies, new policies that try to allow access to a subspace might not have any effect.
In practice, it is likely that a space will be subject to many policies and it can even inherit policies from enclosing spaces, 
which will make it challenging for most users to understand why their new allow policy is not effectual.
Using VACMaps's automated reasoning we can detect these ineffective policies so that the user will be prompted to review and
modify the deny policy in order for the new allow policy to work as expected.


\section{Related Work}
\label{sec:related-work}

This section highlights related work across security and privacy, computer vision, and formal methods.

\subsection{Security and Privacy Approaches to Access Control}

\label{subsec:related-privacy-challenges-in-arvr}

AR security and privacy concerns are not a new concept in the research literature.
Roesner et al.~\cite{roesner_security_2014} point out that AR
systems enable novel attacks against users and pose new threats to
users' and bystanders' privacy. Similarly, De Guzman et al.~\cite{de_guzman_security_2019} surveys recent
developments in the security and privacy of mixed reality (MR) systems by
classifying them based on a data-centric scheme. Chen et al.~\cite{chen_case_2018}
demonstrates that an attacker could use the video feed and depth information
captured by an AR device to recover victim's password input on a touch screen and proposes designing defenses tailored to this. Lebeck et al.~\cite{lebeck_towards_2018} perform a
qualitative lab study on AR headset users and identifies the \textit{need} for 
access control among users to manage shared physical spaces and virtual content
in those spaces. However, their work stops short of implementing an access control system which our work does.

Roesner et al.~\cite{roesner_world-driven_2014} presents a general framework for controlling access to
a continuous stream of sensor data (video and audio) based on policies specified for
real-world objects. 
Our work mainly addresses an orthogonal problem of \emph{reasoning about} policies themselves, to detect potentially problematic policies.

\subsection{Computer Vision Approaches to Access Control}
\label{subsec:related-privacy-cv}

Recent work in computer vision~\cite{akter_i_2020,hasan_automatically_2020} studies how photos and videos taken in
public places might pose a potential privacy risk to bystanders.
The most relevant work to ours in the vision community is by Templeman et al.~\cite{templeman_placeavoider_2014} who present a technique based on image
classification that identifies where an image was taken and recognizes if this
image concerns sensitive spaces like bathrooms or bedrooms. 
This offers the AR device user the capability to ``blacklist'' these spaces by defining their own access control policies. However, this technique only supports simple policies as a
list for forbidden or blacklisted spaces.

\subsection{Formal Methods and Access Control}
\label{subsec:related-formal-methods}
Although there exist standardized frameworks for access control languages, for
example \texttt{XACML}(\cite{noauthor_extensible_2013}), various domain-specific
languages (along with formalization of their semantics) are often designed
to target different application domains. Guelev at al.~\cite{guelev_model-checking_2004}
presents a loop-free programming language and its formal semantics which
correspond to steps that a user could take to try to gain access to a resource.
The authors then move on to reasoning about whether a program exists to prove
that whether or not a user could eventually get access to a resource.
Hughes and Bultan~\cite{hughes_automated_2008} formalizes access policies written in
\texttt{XACML} format as SAT formulas and invokes a SAT solver to check if
policies satisfy a partial order relationship. Dougherty et al.~\cite{dougherty_specifying_2006}
studies access control policies in a dynamic environment and identifies a few
decidable analyses in first-order temporal logic.
Backes et al.~\cite{backes_semantic-based_2018} formalizes semantics of the Amazon Web
Services (AWS) policy language and introduces an analysis tool that encodes
semantics of access policies associated with AWS resources into SMT formulas.
The tool then invokes Z3~\cite{de_moura_z3_2008} (along with a customized theory
solver for strings) to verify properties of AWS access policies.
In contrast, our work is among the first to apply such access control and formal methods techniques to localization and mapping for AR.


\section{Conclusion and Future Work}
\label{sec:conclusion}

This work presents VACMaps which is an access control system that provides provably correct access decisions for AR localization and mapping using formal methods.
VACMaps enables automated reasoning over access policies by introducing a DSL for policy writers and its corresponding semantics via the SMT encoding of the policies. 
We also show that VACMaps is both fast and scalable, and that it can also be used to audit misconfigurations in access policies.

VACMaps focuses on localization and mapping data, however we expect that similar access control needs will arise for other AR data structures and use cases in the future.
Additionally, we hope to introduce a DSL that allows users to write more robust
access policies against estimation errors that are inherent to localization and mapping
algorithms.


\bibliographystyle{ieeetr}
\bibliography{references.bib}

\end{document}